\documentclass[manuscript,acmsmall,nonacm]{acmart}

\usepackage[dvipsnames]{xcolor}
\usepackage{array}
\usepackage{graphicx}
\usepackage[table,xcdraw]{xcolor}
\usepackage{colortbl}
\usepackage{longtable}
\usepackage{subcaption}
\usepackage{float}
\usepackage{longtable}
\usepackage{booktabs}
\usepackage{xcolor}
\usepackage{tcolorbox}
\usepackage{listings}
\usepackage{multirow}
\usepackage{makecell}
\usepackage{xspace}
\usepackage{booktabs}
\usepackage{multirow}
\usepackage{tcolorbox}
\tcbuselibrary{breakable}
\AtBeginDocument{%
  }

\setcopyright{none}
\renewcommand\footnotetextcopyrightpermission[1]{}
\newcommand{\revision}[1]{\textcolor{black}{#1}}

\newcommand{\citex}[1]{\citeauthor{#1}~(\citeyear{#1})}
\newcommand{\xhdr}[1]{\vspace{1.5mm}\noindent\textbf{#1}}

\begin{document}

%%
%% The "title" command has an optional parameter,
%% allowing the author to define a "short title" to be used in page headers.

\title{AI Assistance for Discretionary Work: Increasing Feedback Provision in Higher Education}

%%
%% By default, the full list of authors will be used in the page
%% headers. Often, this list is too long, and will overlap
%% other information printed in the page headers. This command allows
%% the author to define a more concise list
%% of authors' names for this purpose.
\author{Romina Mahinpei}
\email{rmahinpei@princeton.edu}
\affiliation{%
  \institution{Princeton University}
  \country{USA}
}

\author{Victoria Dean}
\email{vdean@princeton.edu}
\affiliation{%
  \institution{Princeton University}
  \country{USA}
}

\author{Ruth Fong}
\email{ruthfong@princeton.edu}
\affiliation{%
  \institution{Princeton University}
  \country{USA}
}

\author{Lydia T.~Liu}
\email{ltliu@princeton.edu}
\affiliation{%
  \institution{Princeton University}
  \country{USA}
}

\author{Manoel Horta Ribeiro}
\email{manoel@cs.princeton.edu}
\affiliation{%
  \institution{Princeton University}
  \country{USA}
}

\renewcommand{\shortauthors}{Mahinpei et al.}



%%
%% Keywords. The author(s) should pick words that accurately describe
%% the work being presented. Separate the keywords with commas.

% \received{20 February 2007}
% \received[revised]{12 March 2009}
% \received[accepted]{5 June 2009}

%%
%% This command processes the author and affiliation and title
%% information and builds the first part of the formatted document.

\begin{abstract}
\revision{AI assistance can increase the efficiency and accuracy of human work, but can it also encourage people to undertake valuable tasks they might otherwise skip? Here, we investigate whether} AI assistance can improve participation in \textit{discretionary but beneficial work} 
in the context of personalized feedback provision in higher education, a pedagogically valuable but often optional practice. We conduct a mixed-methods study combining a randomized field experiment and qualitative interviews in an \revision{upper}-level machine learning course with $n$=11 teaching assistants (TAs) and $n$=88 students. Student submissions were randomly assigned to either (1) a treatment condition where TAs received AI-assisted feedback drafts after grading or (2) a control condition without drafts. 
We find that AI-assisted feedback significantly increases feedback provision (+10.8 percentage points, SE=1.1, $p<0.001$) and feedback length (+39.8 chars, SE=3.45, $p<0.001$) \revision{while also maintaining student-rated usefulness and expert-rated feedback quality among the feedback instances for which these outcomes were observed. 
While time spent conditional on providing feedback did not significantly differ between conditions,  total time spent in the feedback stage increased in the treatment condition (10.55 seconds, SE=4.19, $p=.012$).} Qualitative findings suggest that AI-assisted drafts function as editable scaffolds that lower barriers to initiating feedback rather than reducing overall labor. 
\revision{Together, our findings show that AI assistance can increase participation in discretionary communication without necessarily reducing the labor involved or eliminating the effort required to sustain it.}

\vspace{3mm}
\end{abstract}

\begin{teaserfigure}
\vspace{7mm}
    \centering \includegraphics[width=0.875\linewidth]{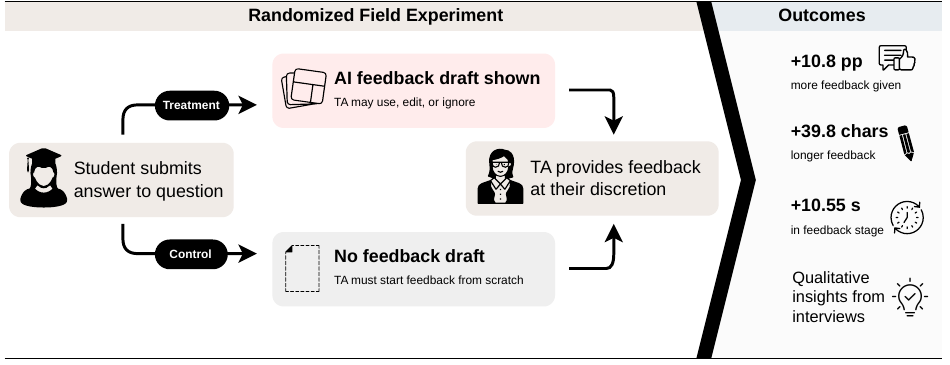}
    \Description{}
    \caption{\textbf{Study overview and main findings}. After grading each question submission, teaching assistants (TAs) are randomly assigned to receive either an AI-assisted feedback draft (treatment) or no feedback draft (control). We find that AI-assisted feedback increased feedback provision and length while also maintaining student-rated usefulness and expert-rated quality. However, AI assistance did not reduce the time spent providing feedback and even increased total time spent in the feedback stage.}
    \label{fig:studies_overview}
\end{teaserfigure}

\maketitle

\section{Introduction}

% discretionary work
AI systems are increasingly integrated into human workflows, shaping how people access information, make decisions, and complete tasks~\cite{Parasuraman00,Amershi19,Shneiderman20}. While some work envisions AI systems as \textit{autonomous agents} that independently act on users' behalf~\cite{Ghai24,Meyer24}, a growing body of research instead positions them as \textit{assistants}: systems that guide human behavior and decision-making by generating intermediate artifacts, such as hints, templates, or drafts, while leaving final decisions to users~\cite{Amershi19,Cui25,Peng23,Zoller25}. Prior work has shown that such AI-assisted workflows can improve task performance in domains including coding~\cite{Ziegler24,Peng23}, writing~\cite{Cui25,Noy23}, and clinical diagnosis~\cite{Zhang24,Zoller25}. However, these studies largely focus on settings in which users are \textit{required} to complete a task, such as engineers completing assigned coding tasks or physicians producing diagnoses. Yet, many socially valuable practices are discretionary: people may intend to mentor, document, review, or provide feedback, yet fail to do so amid competing demands~\cite{organ1988organizational,sheeran2016intention}. \textit{In these settings, the central challenge is not only how well a task is performed, but whether it is performed at all.}

% personalized feedback in higher ed
Personalized feedback in higher education is a prominent example of discretionary work: it is pedagogically valuable, but often under-provided because it is labor-intensive and difficult to scale~\cite{Doe2013,Henderson19,Mahinpei26}. \revision{It is also a form of communication between TAs and students in which TAs translate their evaluation of student work into guidance} that can improve student learning, motivation, and experience~\cite{hattie2007power,shute2008focus,Cambre18,Ahmed2024,Faulconer22,Mandouit23}.
However, providing useful feedback requires TAs to interpret a student’s response, identify misconceptions or opportunities for improvement, and articulate guidance that is specific, constructive, and actionable~\cite{Doe2013,Henderson19,Mahinpei26}. This labor-intensive process often competes with grading workload and other instructional responsibilities.
As a result, feedback is often provided selectively rather than uniformly: during grading, TAs decide whether a response would benefit from feedback and whether they have the time and resources to provide it well. 

The importance of feedback in education has motivated growing interest in using AI to generate personalized feedback at scale, with prior work suggesting that AI-generated feedback can support student learning and improve feedback experiences~\cite{Meyer24,Kinder25,Jacobsen25,Usher25}. In educational settings, however, fully autonomous feedback generation raises concerns about factual accuracy, pedagogical appropriateness, and the risk of giving students misleading guidance~\cite{Ji2023,Wang2024}. For this reason, recent work has explored AI-assisted feedback workflows in which AI systems generate personalized drafts that TAs can review, revise, and choose whether to send. Existing studies suggest that such workflows can improve feedback quality, reduce instructor workload, or support student learning~\cite{Pahi24,Gurung25,Thomas24,Lu26,Wang25}. However, this work largely abstracts away the discretionary aspect of feedback provision. Graders \textit{were required} to provide feedback on every submission, meaning these evaluations measure whether AI improves feedback once it is required, not whether AI changes the decision to provide feedback in the first place. We therefore ask: \textit{Can AI assistance increase feedback provision when giving feedback is optional rather than required?}

\xhdr{Present work.} We address these questions in a real-world higher-education course where feedback provision is discretionary and historically rare. Specifically, we structure our investigation around three research questions:

\begin{itemize}
    \item \textbf{RQ1 -- Feedback Provision Behavior}: How do AI-assisted feedback drafts influence TAs’ feedback provision behavior (e.g., whether feedback is given, its length, and time spent) when giving feedback is discretionary?
    \item \textbf{RQ2 -- Usage Patterns}: How do TAs perceive, use, and adapt the AI-assisted feedback drafts (e.g., usage rate, TA perceived usefulness, similarity between draft and final feedback)?
    \item \textbf{RQ3 -- Downstream Experiences}: How do AI-assisted feedback drafts affect \revision{ quality} and the students' experience of the feedback ultimately provided?
\end{itemize}

We address these questions through a mixed-methods study combining a randomized field experiment with qualitative interviews (see Figure~\ref{fig:studies_overview}). We develop a lightweight, LLM-backed Chrome extension that surfaces personalized feedback drafts directly within TAs' native grading platform. Student submissions throughout the semester were randomly assigned at the question level to either a treatment condition, in which TAs are shown an AI-assisted feedback draft after grading, or a control condition, in which no draft is shown. TAs remain fully in control, being able to use, edit, or ignore each draft at their discretion \revision{and having the final verdict on the feedback that gets communicated to students}. We combine behavioral logs, surveys, and semi-structured interviews to examine how AI-assisted feedback drafts shape the provision of feedback, TA usage, and downstream student experiences \revision{and how these patterns evolve over the semester}.

We find that AI assistance increases discretionary feedback provision by +10.81 percentage points (SE = 1.10, $p < .001$) and feedback length by +39.79 characters (SE = 3.45, $p < .001$), \revision{while not significantly reducing time conditional on providing feedback or time per character}. \revision{At the same time, total time spent in the feedback stage is significantly greater in the treatment condition (10.55 seconds, SE=4.19, $p=.012$), suggesting that AI assistance increases participation without necessarily reducing the human labor involved in producing feedback. Further, the increase in feedback provision between conditions persists across assignments even as absolute feedback provision declines over time, suggesting that supporting initial participation in discretionary work is distinct from sustaining that work over time~\textbf{(RQ1)}.} 

Interviews with TAs further clarify how the work involved in feedback provision is redistributed: drafts function as personalized starting points that help TAs begin writing, articulate comments they already had in mind, verify their interpretation of student responses, and refine the tone or specificity of their guidance. TAs therefore treat AI-generated drafts as editable intermediate artifacts rather than treating them as final outputs \revision{and remain responsible for the work of reviewing, verifying, and editing what is ultimately communicated to students~\textbf{(RQ2)}.} \revision{Further, we find no evidence of reduced student-rated usefulness among rated feedback instances (-0.01, SE = 0.06, $p = 0.88$), while  expert evaluators generally prefer AI-assisted feedback because of its greater specificity~\textbf{(RQ3)}}. \revision{Together, our findings show that AI assistance can increase participation in discretionary communication without simply automating it: students receive more feedback while TAs retain the labor and responsibility of verifying and finalizing that communication}.

\xhdr{Contributions}. Our work makes three primary contributions:

\begin{enumerate}

    \item \textbf{Participation in Discretionary Work}: We show that AI assistance can increase whether optional but beneficial communication between TAs and students happens at all, extending CSCW/HCI research on AI-assisted workflows from improving required tasks to shaping participation in discretionary ones.

    \item \textbf{Redistribution of TA Labor}: We show that increased participation does not necessarily imply reduced human labor: AI produces an intermediate artifact while TAs retain the work of verifying and finalizing the feedback that is ultimately communicated to students.

    \item \textbf{Initiating Versus Sustaining Participation}: We show that AI assistance can continue to increase feedback provision even as absolute provision and draft usage both decline over time, differentiating between supporting the initiation of discretionary work and sustaining participation over time.
\end{enumerate}

Although we study feedback provision in higher education, the broader challenge extends beyond classrooms. Many socially valuable practices, including mentoring, documentation, reviewing, and maintenance work, depend on voluntary effort and are often under-provided. Our findings suggest that AI assistance may be most valuable in these settings not simply as a productivity tool, but as an intervention that changes whether beneficial work is initiated in the first place.

\section{Related Work}
We situate our work in a long CSCW/HCI tradition concerned with how computational systems redistribute initiative, effort, and accountability in human work. While recent generative AI systems have renewed interest in AI-assisted workflows, the design questions are longstanding: when should systems act autonomously, when should they support human judgment, and how do such systems reshape the often-invisible labor on which organizations depend? 
We review work on human-centered AI; CSCW research on invisible, discretionary, and articulation work; and educational research on personalized and AI-supported feedback. 
Across these literatures, we find that existing work primarily asks whether AI improves the quality or efficiency of required tasks, whereas we ask whether AI assistance can increase participation in valuable, optional work that is often skipped.

\subsection{AI Within Human Workflows}

AI systems have long raised questions about how initiative, judgment, and responsibility are distributed between humans and machines~\cite{suchman1987plans,Parasuraman00,Horvitz99,Shneiderman20}. CSCW/HCI research has examined these questions through critiques of plan-based models of action~\cite{suchman1987plans}, mixed-initiative systems that combine automation with direct human control~\cite{Horvitz99}, and analyses of collaborative systems in which costs and benefits are unevenly distributed among actors~\cite{grudin1988cscw}. Building on this lineage, we distinguish between AI systems that act as autonomous agents and AI systems that act as assistants embedded within human workflows \revision{and then discuss how AI has been integrated into collaborative settings involving multiple people}.

\subsubsection{\textbf{AI as Agents}}
AI systems are increasingly integrated into human workflows, shaping how users access information, make decisions, and complete tasks. One line of work in this space positions AI systems as \textit{autonomous agents} that complete tasks on behalf of users, with the AI generating full outputs, such as code, predictions, or feedback, with limited human oversight of the generated content~\cite{Parasuraman00,Ghai24,Meyer24}. 
This paradigm builds on a long tradition of research on automation, which examines how tasks can be delegated to computational systems across varying levels of human control~\cite{Parasuraman00}. It also echoes longstanding HCI debates about machine agency and human action. Classic work on situated action argues that computational systems often fail when they treat work as the execution of predefined plans rather than as situated, context-dependent practice~\cite{suchman1987plans}. 
Most recently, agentic systems have been explored across various domains, including software engineering, where AI systems independently generate code~\cite{Ghai24}, and education, where AI systems automatically generate feedback for students~\cite{Meyer24}. While such systems can improve efficiency and accuracy, they often limit opportunities for human involvement and raise concerns about reliability, transparency, and over-reliance~\cite {lee2004trust,endsley2018automation}.

\subsubsection{\textbf{AI as Assistants}} 
An alternative line of work positions AI systems as \textit{assistants} that support, rather than replace, human behavior and decision-making~\cite{Amershi19,Shneiderman20}. This framing builds on longstanding research in mixed-initiative interaction and interactive machine learning, which examines how computational systems can share initiative with users and how people can guide, correct, and incorporate model outputs into their own work~\cite{Horvitz99,amershi2014power}. Rather than autonomously completing tasks on behalf of users, these systems generate intermediate artifacts, such as hints, suggestions, drafts, or templates, that humans can selectively incorporate into their workflows. This paradigm aligns with broader visions of human-centered AI that emphasize complementarity between humans and AI systems, aiming to provide high levels of computational support while preserving human oversight, agency, and control over final outcomes~\cite{Shneiderman20,Amershi19}.
At the same time, prior HCI work shows that AI-assisted workflows are difficult to design because AI systems are uncertain, adaptive, and capable of producing complex outputs that users must interpret and evaluate~\cite{yang2020re}.

Whereas agentic systems are often evaluated for their ability to complete tasks or optimize task outcomes, AI-assisted systems are commonly designed to shape how people engage with tasks by providing structured starting points. Prior work shows that such systems can improve performance and productivity across domains, including writing~\cite{Cui25,Noy23}, software development~\cite{Peng23}, and clinical decision-making~\cite{Zhang24,Zoller25}. For example, studies of generative AI support for professional writing and software development typically evaluate outcomes such as completion time, output quality, or productivity~\cite{Noy23,Peng23}. \revision{Taken together, this line of work highlights how AI assistance can alter how people perform work while maintaining human involvement in producing the final output.}
% ORIGINAL ENDING
% However, recent evaluations of AI-assisted workflows largely focus on settings in which task completion is \textit{required}, with outcomes centered on performance, efficiency, or accuracy. In contrast, many socially valuable practices, such as mentoring, documentation, and feedback provision, are \textit{discretionary} and often under-provided. In these settings, the key challenge is not only how well a task is performed, but whether it is performed at all. We therefore study whether AI assistance can increase participation in optional tasks that are often skipped.

\subsubsection{\textbf{\revision{AI Assistance in Collaborative Settings}}} 
\revision{AI assistance has also been explored by the CSCW/HCI community within collaborative and communicative workflows involving multiple people, where AI support can mediate how people communicate, coordinate, and contribute to shared work. Prior work has shown both the potential as well as limitations of introducing AI into these settings~\cite{Mieczkowski21AIMC,Xiao24AIAssistedEditing,Zhang22AITranslationsMeetings,Chen25LLMsMappingMeetings,Asthana25LLMRecaps}. For example, in the context of group meetings, AI support has been found helpful for translating pre-meeting discussions for multilingual groups~\cite{Zhang22AITranslationsMeetings}, creating non-linear representations of ideas discussed~\cite{Chen25LLMsMappingMeetings}, and generating meeting recaps~\cite{Asthana25LLMRecaps}. At the same time, other work has found that AI-mediated communication can exhibit positivity bias and undermine certain aspects of interpersonal perception~\cite{Mieczkowski21AIMC} and that AI-assisted editing does not exempt non-native speakers in co-writing settings from being disadvantaged~\cite{Xiao24AIAssistedEditing}. Together, this work highlights how the consequences of AI assistance can extend beyond individual task performance to the interactions and relationships through which the work is achieved.}

% \revision{Across these settings, AI assistance has been primarily studied in contexts in which task completion is \textit{required}, with outcomes centered on performance, efficiency, or accuracy. In contrast, many socially valuable practices, such as mentoring, documentation, and feedback provision, are \textit{discretionary} and often under-provided. In these settings, the key challenge is not only whether AI can support how well a task is performed, but whether it can change participation in work that might otherwise not be performed at all. We therefore study whether AI assistance can increase participation in optional communicative work, focusing on TAs' provision of discretionary feedback to their students as a form of communication.}

\subsection{\revision{Invisible} and Discretionary Work}
Many socially valuable forms of work are invisible or discretionary, meaning they are neither formally acknowledged nor strictly required despite their benefits. \revision{CSCW/HCI research has long examined such work through the distinct but overlapping concepts of \textit{articulation} or \textit{invisible} work as well as \textit{voluntary} or \textit{discretionary} work.} 

\subsubsection{\textbf{\revision{Invisible Work}}} Invisible work refers to the often-unseen labor required to coordinate tasks, people, and breakdowns in collaborative settings~\cite{schmidt1992taking}. Similarly, ~\citex{star1999layers} argue that invisibility is not simply a property of work itself, but a social and organizational process through which some forms of labor become unrecognized, unmeasured, or taken for granted. This framing connects to broader labor scholarship showing that conventional notions of work often privilege activities that are formally compensated or directly tied to measurable productivity, while overlooking unpaid, informal, emotional, or supportive labor~\cite{Daniels87, Budd16,Hatton17}.

Across domains, invisible work is often essential despite being undervalued. Prior work has documented these dynamics in care work~\cite{Glenn00}, disability-related labor~\cite{Devault14}, organizational coordination in nursing~\cite{Allen14}, unpaid digital labor online~\cite{Scholz12}, and open-source software ecosystems, where substantial community labor involves moderation, coordination, mentorship, and other non-code contributions~\cite{Meluso25}. These forms of work frequently fall outside formal incentives and expectations. As a result, individuals may deprioritize this work despite recognizing its broader value, particularly when it requires additional time, effort, or investment beyond their primary responsibilities.

\subsubsection{\textbf{\revision{Discretionary Work}}} 
\revision{Discretionary work refers to the voluntary labor performed often as part of sustaining community-led spaces and processes, such as maintaining open-source software, moderating content, or supporting peer review~\cite{Geiger21LabourOSS,Zhang24DSMAdmins,Sun24AIPeerReview}. Similar to invisible work, discretionary work can often go unappreciated and not formally rewarded albeit being potentially more visible (e.g., the outcome of a peer review process is visible to the receiver). For example, \citex{Geiger21LabourOSS} document the broad range of voluntary work performed by maintainers of open-source software and how it goes beyond software code and also includes often invisible interpersonal and organizational work in sustaining such projects. Similarly, \citex{Kriplean08WikipediaWork} have documented the range of volunteer activities involved in maintaining Wikipedia as open knowledge platform and \citex{Zhang24DSMAdmins} have detailed the burdens of administrators and moderators in open-source social media platforms, especially with the growing scale of online interactions.}

\subsubsection{\textbf{\revision{Supporting Invisible and Discretionary Work}}} 
Prior CSCW/HCI research has explored ways to make both invisible and discretionary work more visible and to encourage participation in overlooked activities~\cite{star1999layers}. For example, \citeauthor{Kow18} developed a workplace system that allowed employees to compliment one another for otherwise hidden contributions, finding that increased visibility helped motivate participation in these activities~\cite{Kow18}. Other work has developed tools to quantify otherwise hidden labor, such as the unpaid searching, communication, and payment-management work performed by crowd workers~\cite{toxtli2021quantifying}. In open-source communities, researchers have also argued for recognition and attribution mechanisms for non-code contributions such as moderation, coordination, mentorship, outreach, and community management~\cite{young2021contributions,Meluso25}. 
\revision{Similarly, researchers have explored AI-driven tools for supporting moderation work in social media~\cite{Kuo23FacebookModeration} as well as drafting meta reviews during the peer review process~\cite{Sun24AIPeerReview}, aiming to ease the burden that may be imposed by such voluntary but beneficial work.}
These studies suggest that participation gaps can sometimes be addressed by changing how work is surfaced, recognized, or supported.

% \revision{In this work, we focus on \textit{feedback provision} as a form of discretionary communicative work.} While feedback is critical for learning, providing it is often optional and inconsistently implemented~\cite {Doe2013, Mahinpei26, Henderson19}. We specifically examine whether AI assistance can help address participation gaps in discretionary work by lowering the barriers to providing personalized feedback.

\subsection{AI for Personalized Feedback Provision}
Feedback plays an important role in student learning, motivation, and academic success, but its effects depend on its content, timing, and relationship to students' goals and understanding~\cite{Cambre18, Ahmed2024, Faulconer22, Mandouit23, shute2008focus, hattie2007power}. 
Prior work on formative feedback emphasizes that effective feedback should help learners understand the gap between current and desired performance and support subsequent action, rather than simply evaluate correctness~\cite{hattie2007power,nicol2006formative}.
Thus, personalized feedback is especially valuable because it can tailor guidance to individual students' needs, progress, and understanding~\cite{Narciss14,Paterson20,Fleur23}.
Students also express strong preferences for feedback that is individualized, specific, and actionable~\cite{Paterson20}.

Providing such feedback, however, is difficult to scale in large higher-education courses. Personalized feedback requires instructors and TAs to interpret student work, identify the relevant misconception or opportunity for improvement, and communicate guidance in a way that is both useful and motivating. As a result, feedback is often inconsistent under high workloads and competing responsibilities, especially when giving feedback is discretionary rather than required~\cite{Doe2013,Mahinpei26,Henderson19}. These challenges have motivated growing interest in using AI systems to support feedback provision in higher education.
 
\subsubsection{\textbf{AI-Generated Feedback}}
One line of work explores the use of LLMs to automatically generate feedback directly for students. Several studies suggest that AI-generated feedback can improve student outcomes and experiences. For example,~\citex{Meyer24} shows that GPT-generated feedback improves revision quality, motivation, and affect while~\citex{Kinder25} finds improvements in written justifications among pre-service teachers. Other work suggests that AI-generated feedback can outperform novice human feedback~\cite{Jacobsen25} and may be perceived as higher quality than peer feedback~\cite{Usher25}.
Despite these promising findings, fully automated feedback remains controversial. Some studies find no significant differences in learning outcomes between AI-generated and human-generated feedback~\cite{Kinder25, Escalante23}, and student preferences are often mixed~\cite{Escalante23}. More importantly, prior work highlights reliability concerns, including hallucinations and factual inaccuracies~\cite{Jia24, Wang2024, Milano2023, Yan2024}, as well as pedagogical limitations, such as revealing answers rather than supporting student reasoning~\cite{Wang2024}. These limitations raise concerns about over-reliance and motivate approaches that preserve human oversight in the feedback process.

\subsubsection{\textbf{AI-Assisted Feedback}} 
In response to concerns about fully automated feedback, an emerging line of work explores AI-assisted feedback, in which AI systems support rather than replace instructors and TAs. In these systems, the AI produces suggestions, hints, or feedback templates that TAs can review and refine before sending to students. Prior work suggests that such forms of human–AI collaboration can improve feedback quality and student learning outcomes. For example, studies comparing human–AI tutoring to AI-only tutoring found that human tutors supported by AI systems achieve greater student learning gains than AI tutors alone~\cite{Gurung25, Thomas24}. Similarly,~\citex{Pahi24} found that collaborative TA--AI feedback outperforms both human-only and AI-only feedback, while~\citex{Lu26} shows that AI-mediated feedback improves student revisions when TAs remain in control of whether and how to use AI-generated suggestions.

However, these studies largely evaluate AI-assisted feedback in settings where feedback provision is required. In tutoring experiments~\cite{Gurung25, Thomas24}, tutors are required to provide instruction, and in feedback studies~\cite{Pahi24, Lu26}, 
TAs provide feedback as part of the study workflow. These designs are well-suited to evaluating feedback quality, revision outcomes, or learning gains conditional on the provision of feedback. They leave open the question that arises in many real-world courses: whether AI assistance changes the likelihood that feedback is provided at all. In these settings, feedback provision is often discretionary and infrequent because TAs face time constraints, competing responsibilities, and limited incentives to write individualized comments. This leaves an important open question: \textit{how does AI assistance influence feedback provision when it is optional?}

To address this gap, we study AI-assisted feedback in a higher-education course setting where feedback provision is discretionary. Using a randomized field experiment, we examine how AI assistance affects feedback provision behavior, usage patterns, and downstream student experiences. \revision{Together, our work extends prior CSCW/HCI research on AI-assisted workflows in collaborative and communicative settings by shifting attention from AI support for required tasks to whether AI assistance can increase participation in valuable but optional work, focusing on TAs’ provision of discretionary feedback to students.}

\section{Course Context}
We conducted our study in an \revision{upper-level} undergraduate machine learning (ML) course (``Course X'') at a private R1 university. Course X typically enrolls 130-150 students per semester, including both CS majors and students from other disciplines, and is a core part of the undergraduate curriculum. The course emphasizes conceptual understanding of ML algorithms through written assessments, including two exams and four homework assignments (HW1-HW4). These assessments consist of two question types: (1) \textit{applied questions}, which require applying known methods to concrete problems, and (2) \textit{conceptual questions}, which require reasoning about underlying principles and extending them to new contexts (see Appendix~\ref{app:sample_problems} for a sample problem for each question type).

All written work is manually graded by teaching assistants (TAs), including both \textit{graduate TAs (GTAs)} and \textit{undergraduate TAs (UTAs)}. To ensure grading consistency, each question is assigned to a single TA, who evaluates all submissions using instructor-provided rubrics. Within this structured grading workflow, however, written feedback is optional: TAs may choose whether to leave individualized comments in addition to assigning scores. As a result, feedback has historically been rare in Course X, making it a natural context for studying whether AI assistance can increase participation in a discretionary but pedagogically valuable task.

\section{Formative Study}

\begin{figure*}[t]
\centering
\includegraphics[width=\textwidth]{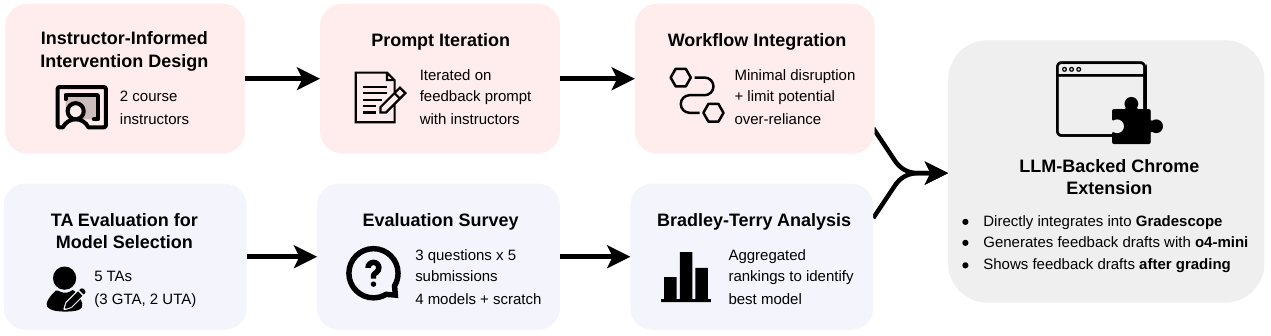}
\caption{\textbf{Overview of the formative study}. Two parallel components informed the design of our AI-assisted feedback intervention: (1) instructor-informed intervention design, which included iterative prompt engineering and workflow design with course instructors, and (2) a TA evaluation for model selection, in which TAs ranked feedback drafts generated by four candidate LLMs and an option to write feedback from scratch.}
\label{fig:formative_study}
\end{figure*}

To inform the design of our feedback generation process and its integration into Course X, we conducted a formative study with instructors and TAs (see Figure~\ref{fig:formative_study}). Our goal was to develop a lightweight intervention that could lower the effort required to provide individualized feedback without disrupting existing grading practices. The formative study consisted of two components.
First, we worked with course instructors to define pedagogical and workflow requirements for AI-assisted feedback. 
Second, we conducted a TA evaluation to compare feedback drafts generated by candidate LLMs and select the model used in the randomized field experiment.  
\revision{Because TA assignments for the targeted course offering were not finalized until a few weeks before the semester, after the intervention design and IRB protocol needed to be finalized, we recruited TAs from previous offerings of the course to incorporate their perspectives through the model selection process}.
All instructors and TAs in the formative study provided consent before participating.

\subsection{Instructor-Informed Intervention Design}
We worked closely with the two instructors of Course X to define requirements for AI-assisted feedback that would be pedagogically appropriate and feasible to integrate into the course's grading workflow. As a starting point, we adapted a feedback prompt from \cite{Mahinpei26arxiv}, which had been shown to produce high-quality feedback drafts in a similar undergraduate-level computer science course. Because feedback in Course X was historically rare, we did not have a dataset of instructor or TA feedback to fine-tune on. Instead, we relied on iterative prompt engineering with instructors to ensure that generated feedback aligned with course conventions and expectations. The final prompt (Appendix~\ref{app:feedback_prompt}) specifies the model’s role and incorporates established feedback design principles \cite{Narciss08}. For each submission, the prompt takes as input: (1) the problem description, (2) the instructor solution, (3) the grading rubric, and (4) the student submission.

These discussions also surfaced key design considerations for integrating the intervention into Course X's grading workflow. First, instructors emphasized \textit{minimizing disruption to existing grading practices}. To address this, we implemented a lightweight Chrome extension that surfaced the feedback drafts directly within the course’s grading platform (see Section~\ref{sec:chrome_extension} for more details). Second, instructors raised \textit{concerns about potential over-reliance on AI-assisted feedback}. To mitigate this risk, we intentionally separated the grading task from feedback provision: TAs first completed grading independently, and only then proceeded to the feedback stage, at which point the feedback drafts were presented (see Section~\ref{sec:experiment_design} for more details).

\subsection{TA Evaluation for Model Selection}
To ensure our feedback drafts would be useful and appropriate, we conducted a TA evaluation to compare models and identify the one that best supports AI-assisted feedback. \revision{We selected four candidate models (o4-mini, DeepSeek-R1, Llama-4-Maverick, Mistral-Large-2411) for this process based on their availability through our institution-managed Azure subscription at the time of the formative study (June–July 2025). We restricted our selection to models available through this service to comply with institutional data-handling policies for student data, such as homework submissions.}

\subsubsection{\textbf{Procedure}}
We constructed a survey using 3 past questions from Course X (2 conceptual, 1 applied), each paired with 5 student submissions (2 with mistakes, 3 without). For each submission, participants were shown the problem statement, instructor solution, and grading rubric, followed by feedback generated by the four candidate models, as well as an option to write feedback from scratch (``Scratch''). Participants were asked to rank the five options and provide qualitative feedback on their preferences. To recruit participants for our study, we sent out email invitations to all individuals who had served as TAs in the past four offerings of Course X. We successfully recruited 5 TAs (3 GTAs, 2 UTAs).

\subsubsection{\textbf{Analysis}}
We aggregated rankings across submissions and estimated latent preference scores using a Bradley-Terry model \footnote{As implemented in the \href{choix}{https://pypi.org/project/choix/} package.}, with 95\% confidence intervals computed via bootstrapping ($n=1{,}000$). In addition, the first author analyzed participants’ written rationales using affinity diagramming and clustered responses to identify recurring themes.

% Specifically, we infer parameters 
% $\mathbf{\theta}=\{
% \theta_{\text{o4-mini}}, 
% \theta_{\text{DeepSeek}},
% \theta_{\text{Llama}},
% \theta_{\text{Mistral}},
% \theta_{\text{Scratch}}
% \}$, such that the probability of a feedback option (e.g., $a$) being preferred over the others (e.g., $\{b,c,d,e\}$) is given by:
% $$
% p(a \succ \{b,c,d,e\}) = \frac{e^{\theta_a}}{
%     e^{\theta_a} + e^{\theta_b}  + e^{\theta_c} + e^{\theta_d} + e^{\theta_e}}. 
% $$

\subsubsection{\textbf{Results}}
The Bradley-Terry analysis indicated that o4-mini was the most preferred model [$\theta_{\text{o4-mini}}$ = 0.446; 95\% CI: (-0.002, 0.906)], followed by DeepSeek [$\theta_{\text{DeepSeek}}$ = 0.254; 95\% CI: (-0.254, 0.715)], Mistral [$\theta_{\text{Mistral}}$ = 0.251; 95\% CI: (-0.264, 0.720)], and Llama [$\theta_{\text{Llama}}$ = 0.123; 95\% CI: (-0.408, 0.610)], with the Scratch option receiving the lowest preference [$\theta_{\text{Scratch}}$ = -1.094; 95\% CI: (-2.310, -0.383)]. Consistent with these results, o4-mini also received the highest number of first-place rankings, leading us to select it as the model for our randomized field experiment.

Qualitative analysis revealed several themes that informed the design of the feedback drafts. First, TAs preferred feedback that was \textit{specific and actionable}, particularly when it clearly identified correct and incorrect parts of a submission and provided guidance without fully revealing solutions. 
% As one participant noted, ``I like the responses that say explicitly what the student had that was incorrect" while another similarly stated, ``Good to have templates that hint at should be fixed in the submission", suggesting that participants liked when the templates identified which parts of a submission were correct or incorrect while providing hints that guided students.
Second, TAs emphasized the importance of \textit{conciseness}, especially for correct submissions, where brief acknowledgment and reinforcement were preferred over detailed explanations. 
% As one participant noted, ``I think this is just a typo and doesn't need a super elaborate explanation” while another similarly stated that ``I prefer shorter templates for fully correct submissions as well as the ones that included positive reinforcement and mentioned what the submission had done well”.
Third, participants highlighted \textit{stylistic considerations}, such as directly addressing students (e.g., using “you”), to improve clarity and tone. Finally, TAs raised concerns about \textit{over-reliance on AI-assisted feedback}, noting that graders might defer to the feedback drafts without sufficient verification.
% ``I think it might lead TAs to rely on the LLM response even if it's not correct. I also think it reduces the need for the TA/grader to really understand the material".

Altogether, these findings informed both our model selection and prompt design. We selected o4-mini as the underlying model and incorporated additional instructions to generate concise, actionable, and appropriately styled feedback. In addition, concerns about over-reliance reinforced our decision to separate grading from feedback provision and introduce AI-assisted feedback drafts only after grading was completed, as was also requested by the course instructors.

\section{Randomized Field Experiment}

Using insights from our formative study, we conducted a pre-registered\footnote{Pre-Registration Link: \url{https://osf.io/9r2ka/overview}}, semester-long randomized field experiment in Course X to evaluate the impact of AI assistance on feedback provision behavior, TA usage patterns, and downstream student experiences. The intervention was integrated into the course's normal grading workflow through a lightweight Chrome extension that surfaced AI-assisted feedback drafts after TAs completed grading. The study was reviewed and approved by our institution's IRB (Number: 18396), and all participating TAs and students provided informed consent.

\subsection{Participants}
Our study was conducted during the Fall 2025 offering of Course X. All $n=11$ TAs in the course (7 GTAs and 4 UTAs) \revision{who had grading responsibilities} opted into the study. TAs were informed that some submissions would be accompanied by AI-assisted feedback drafts that they could choose to use as part of an intervention integrated into their grading workflow. Students were similarly informed that some of the feedback they received might be AI-assisted. A total of $n=88$ students opted to receive potential AI-assisted feedback, complete optional post-homework surveys evaluating the usefulness of the feedback they received, and participate in optional end-of-semester interviews. TAs received a \$50 gift card upon completion of the study. Students were entered into raffles for \$25 gift cards for each submission of a post-homework survey.

\begin{figure}[t]
\label{fig:chrome_extension}
\centering
\begin{subfigure}[t]{0.3\textwidth}
    \centering \includegraphics[height=7cm]{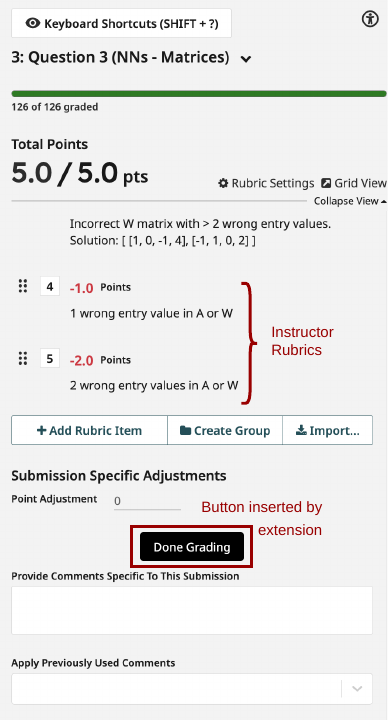}
    \caption{\textbf{Grading interface} with \texttt{Done Grading} button inserted by our Chrome extension}
    \label{subfig:gradescope}
\end{subfigure}
\hfill
\begin{subfigure}[t]{0.3\textwidth}
    \centering \includegraphics[height=7cm]{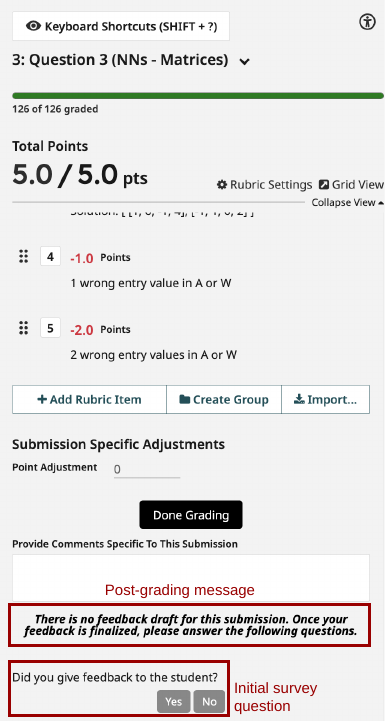}
    \caption{\textbf{Control condition} without AI-assisted feedback draft}
    \label{subfig:control}
\end{subfigure}
\hfill
\begin{subfigure}[t]{0.3\textwidth}
    \centering \includegraphics[height=7cm]{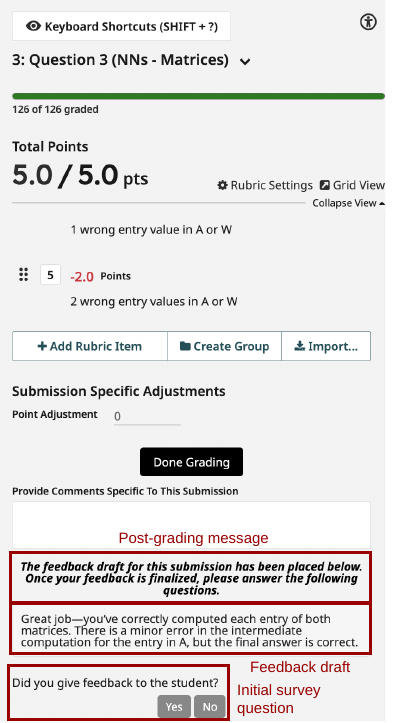}
    \caption{\textbf{Treatment condition} with AI-assisted feedback draft}
    \label{subfig:treatment}
\end{subfigure}
\caption{\textbf{Integration of AI-assisted feedback drafts into the grading workflow}. We develop a lightweight, LLM-backed Chrome extension that surfaces feedback drafted after grading via a “Done Grading” button (A). In the control condition, TAs proceed without assistance (B), whereas in the treatment condition, they are shown a personalized AI-assisted feedback draft (C). Both conditions included a post-grading message and the same initial survey question to \revision{reduce potential differences in nudging between conditions.}}
\label{fig:chrome_extension}
\end{figure}

\subsection{Chrome Extension Design}
\label{sec:chrome_extension}

To minimize disruption to existing workflows, we implemented a lightweight, LLM-backed Chrome extension that we integrated into the course's grading platform (Gradescope). The extension, powered by o4-mini, generated personalized feedback drafts for each submission using the prompt from our formative study, \revision{with example drafts in Appendix~\ref{app:feedback_drafts}}. After grading a submission, TAs click a \texttt{Done Grading} button inserted by the extension. At this point, the system displays a post-grading message indicating whether an AI-assisted feedback draft is available.  In the treatment condition, this message is followed by an AI-assisted feedback draft; in the control condition, no feedback draft is shown (see Figure~\ref{fig:chrome_extension}). For treatment submissions, TAs can use the feedback draft as-is, edit it, or ignore it when deciding whether to provide feedback. This design ensured that AI assistance was introduced only after grading was complete, preserving TAs' independent evaluation of student work and reducing the potential for over-reliance, as requested during the formative study.

As part of the feedback process, TAs also completed a brief \textit{in situ} survey embedded into Gradescope by the extension, indicating whether they provided feedback. For submissions in the treatment condition, TAs were additionally asked (1) whether they used the feedback draft, (2) how helpful they found the draft (1--5 Likert scale), and (3) whether they would consider revising the assigned grade after viewing the draft. We also logged behavioral data, such as the final feedback that was provided, the length of the provided feedback, and the time spent in the feedback stage (measured from clicking \texttt{Done Grading} to the start of the survey). 

Importantly, both conditions included a post-grading message and the same initial survey question about whether feedback was provided. \revision{This was intended to reduce potential differences in nudging between conditions, although the messages differed in their wording to reflect whether an AI-assisted draft was available or not (see Section~\ref{sec:limitations} for a discussion of this limitation).} Lastly, to validate our Chrome extension and integration plan, we conducted a pilot during a practice assignment (HW0) administered prior to the study, which revealed no design issues requiring changes to the system or experiment.

\subsection{Experiment Design}
\label{sec:experiment_design}
\begin{figure*}[t]
\centering
\includegraphics[width=0.8\textwidth]{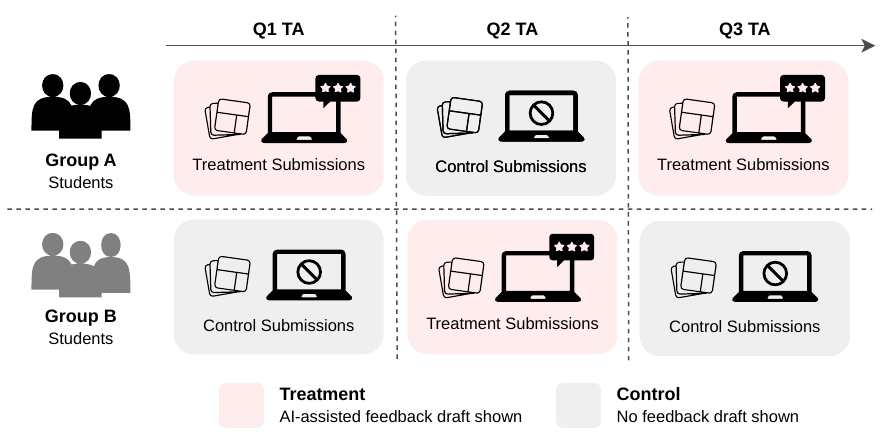}
\caption{\textbf{Question-level randomization with rotating assignment}. Students were randomly assigned to Group A or B at the start of each assignment. Treatment and control conditions alternated across questions within an assignment: for Question 1, Group A received treatment and Group B control; for Question 2, the assignment reversed; and this rotation continued for subsequent questions. Each question was graded by a single TA so that all submissions for a given question were evaluated by the same grader.}
\label{fig:experiment_design}
\end{figure*}

\begin{table*}[b]
\centering
\footnotesize
\caption{\textbf{Outcome measures and sample sizes.} We report all measures used in our analysis, along with their effective sample sizes at the course level. Sample sizes vary across outcomes because some measures are conditional on prior actions (e.g., feedback time is only defined when feedback is provided; feedback rating is only defined when a student rated feedback when it was provided), and feedback draft-related measures are only defined in the treatment condition. \revision{Feedback stage timing data are missing for some submissions because some TAs initially used an earlier version of the extension from the HW0 pilot that did not record timing information; this issue was corrected once detected.}}
\label{tab:measures}
\begin{tabular}{p{2.8cm} p{8.4cm} r}
\toprule
\textbf{Measure} & \textbf{Description} & \textbf{$N$ (Ctrl / Trt)} \\
\midrule
Feedback Provision & Whether a TA provided feedback on a submission & 1{,}418 / 1{,}410 \\

Feedback Length & Length of feedback in characters (0 if no feedback),  & 1{,}418 / 1{,}410  \\

\revision{Feedback Stage Time} & \revision{Time spent in the feedback stage (seconds)} & \revision{1{,}234 / 1{,}227} \\

\revision{Feedback Time} & \revision{Time spent in the feedback stage when feedback was given (seconds)} & \revision{276 / 388} \\

\revision{Adjusted Feedback Time} & Time spent per character of feedback (seconds/character) & 276 / 388 \\

Feedback Rating & Student-rated usefulness of received feedback (1--5 Likert scale) & 214 / 254 \\

Draft Usage & Whether a feedback draft was used (as-is or modified) in the final feedback & -- / 458 \\

Draft Rating & TA-rated usefulness of the feedback draft (1--5 Likert scale)  & -- / 1{,}410 \\

Similarity Score  & Similarity between draft and final feedback, measured by ROUGE-L F1 & -- / 432 \\

Grade Reconsideration & Whether a TA would change the assigned grade after viewing the draft & -- / 1{,}410\\
\bottomrule
\end{tabular}
\end{table*}

\subsubsection{\textbf{Field Experiment}} We conducted a semester-long randomized field experiment across the four written assignments (HW1–HW4) in Course X. As enforced by the course instructors, each assignment question was graded by a single TA, such that all submissions for a given question were evaluated by the same grader. For each assignment, students were randomly partitioned into two groups (A and B). Randomization occurred at the question submission level using a rotating assignment scheme: for the first question, submissions from Group A were assigned to the treatment condition and submissions from Group B to control; for the second question, the assignment was reversed, and this pattern continued across subsequent questions (see Figure~\ref{fig:experiment_design}). In the \textbf{treatment condition}, TAs received an AI-assisted feedback draft after completing grading. TAs could use the feedback draft as-is, edit it, or ignore it. In the \textbf{control condition}, no feedback draft was shown, though TAs could still provide feedback independently. Importantly, feedback provision was optional and left to the TA's discretion in both conditions, allowing us to directly examine whether AI assistance influences participation in discretionary feedback provision.

We examined how AI assistance influences feedback-provision behavior, usage patterns, and downstream student experiences through the outcomes reported in Table~\ref{tab:measures}. Behavioral measures (e.g., feedback provision, length, and timing) were logged automatically, whereas perception-based measures (e.g., feedback and draft ratings) were collected via surveys. Draft usage was defined as whether the AI-assisted feedback draft was used as-is or with modifications when the TA chose to provide feedback. Similarity scores were computed whenever both the generated draft and the final submitted feedback were available, regardless of self-reported draft usage, because textual similarity may capture implicit or unreported reliance on AI-assisted drafts. We specifically computed ROUGE-L F1 as our metric for similarity score, which measures the longest common subsequence shared between two texts, with values closer to 1 indicating greater textual overlap between the generated draft and final submitted feedback and values closer to 0 indicating greater divergence between the two texts~\cite{Lin04}. We use this metric as a proxy for the extent to which TAs reused or adapted the AI-assisted feedback drafts. 

\revision{\subsubsection{\textbf{Feedback Quality Evaluation}} We recruited three course experts (two instructors and one head graduate TA with a supervisory role) who had not participated in the field experiment to evaluate the quality of the final feedback provided for a subset of control and treatment submissions. We matched control and treatment submissions without replacement at the question-submission level, requiring both submissions in each pair to have received some feedback and the exact same score. This resulted in 115 matched pairs across all homework assignments. For each pair, experts were provided with the original question, grading rubrics, student submissions, and the final score, but were blinded to treatment assignment. The presentation order of control and treatment feedback within each pair was randomized.}

\revision{We first conducted a calibration exercise in which all three experts independently evaluated the same 10 pairs and subsequently discussed their evaluations. For each pair, experts indicated (1) which feedback response they preferred or if the preference was about the same; (2) whether the preference was attributed to any subset of correctness, specificity, actionability, pedagogical appropriateness, or another factor they could specify; and (3) whether either response contained issues such as answer leakage, hallucinations, or other concerns they could specify.}

\revision{During calibration, at least two out of the three experts agreed on their preference for 9 of the 10 pairs. Disagreements primarily involved distinguishing a slight preference from no preference. Based on this discussion, we revised the evaluation instrument to distinguish between mild and strong preferences in the first question. We then randomly distributed the remaining 105 pairs among the three experts, with each independently evaluating 35 distinct pairs using the revised instrument provided in Appendix~\ref{app:feedback_quality_eval}.}

\subsubsection{\textbf{Semi-Structured Interviews}} We lastly conducted semi-structured interviews with all TAs ($N=11$) and a subset of students ($N=9$) at the end of the semester to understand their respective experiences with AI-assisted feedback. All interviews were conducted and transcribed via Zoom, and the first author verified the transcripts for accuracy. Interviews lasted 30 minutes on average. We provide all interview questions in Appendix~\ref{app:interview_questions}. 

\subsection{Analysis Approach}

\subsubsection{\textbf{Estimation of Treatment Effects}}
We pool observations at the question-submission level across all four homework assignments. Because our intervention was randomized within each homework question, our primary analyses estimate treatment effects by comparing treated and control submissions within the same homework question. Specifically, for outcomes observed in both conditions, we estimate linear regression models of the form:

\[
Y_{ihq} = \alpha_{hq} + \beta T_{ihq} + \varepsilon_{ihq},
\]

where \(Y_{ihq}\) denotes the outcome for student \(i\) on question \(q\) of homework \(h\), \(T_{ihq}\) indicates assignment to the AI-assisted feedback draft condition, and \(\alpha_{hq}\) denotes homework question fixed effects. These fixed effects absorb all factors constant within a given homework question, including the assigned TA, grading rubric, and question characteristics. The coefficient \(\beta\) therefore estimates the \textit{average within-question effect of AI assistance}. We report cluster-robust standard errors clustered at the student level to account for repeated observations from the same student across questions and homework assignments.

For feedback provision, we use a linear probability model so that treatment effects can be interpreted directly as percentage-point changes. For feedback length, our specification assigns submissions without feedback a value of zero, capturing the total amount of feedback made available to students. 
\revision{For feedback time and adjusted feedback time}, we restrict the analysis to submissions that received feedback, because feedback provision is itself affected by treatment. \revision{To complement these conditional analyses, we also report feedback stage time across all submissions, without this restriction}. Similarly, for feedback ratings, we restrict the analysis to submissions that received feedback as well as a student rating. These conditional estimates are interpreted as descriptive evidence about the feedback that was produced, rather than as standalone causal effects.

\subsubsection{\textbf{Descriptive and Subgroup Analyses}} 
Outcomes observed only in the treatment condition or contingent on specific actions are analyzed descriptively. These include draft usage, draft rating, similarity score, and grade reconsideration. We report means and standard errors overall, by homework, and across relevant subgroups (TA role, question type, and presence of mistakes). \revision{We selected these three categories as they represent the available submission- and grader-level metadata for examining variation across grading contexts. We also provide the corresponding subgroup sample sizes in Appendix~\ref{app:subgroup_sample_size} (Table~\ref{tab:subgroup_sizes})}. These subgroup analyses are interpreted descriptively and are not used for causal identification. We additionally report overall and homework-level summaries, as well as stratified summaries by relevant subgroups for outcomes observed in both experimental conditions in the Appendix~\ref{app:summary_metrics}. 

\subsubsection{\textbf{\revision{Feedback Quality Analysis}}}
\revision{We restricted our analysis to the 105 post-calibration pairs, with 35 pairs independently evaluated by each expert. We first computed the percentage of pairs for which experts preferred treatment feedback, preferred control feedback, or expressed no preference. Among pairs with a preference, we calculated the proportion of mild versus strong preferences separately for treatment and control. We also calculated the proportion for which each quality factor was selected as a reason for the preference separately for treatment and control. Because experts could select multiple factors, these percentages do not necessarily sum to 100\%.}

\subsubsection{\textbf{Thematic Analysis}}
We conducted an inductive thematic analysis of our interview data. The first author conducted an initial round of open coding on a subset of transcripts (two TA and two student interviews), using individual question responses as the unit of analysis to develop an initial codebook. The codes were iteratively refined through discussions with the research team, resulting in a finalized codebook, which the first author then used to recode the full set of transcripts. The research team then synthesized the coded data into higher-level themes that characterize TA and student experiences with AI-assisted feedback.

\section{Findings}
We present our findings on how AI assistance influences feedback provision behavior (RQ1), usage patterns (RQ2), and downstream user experiences (RQ3). \revision{Within each research question, we first report the relevant quantitative measures and then use our interview findings to contextualize and explain the observed patterns.} To support reproducibility, we also provide our dataset and analysis scripts in our repository.
\footnote{Repository Link: \url{https://github.com/humans-and-machines/ai-feedback-provision}}

\begin{table}[b]
\centering
\small
\caption{\textbf{Regression results for outcomes observed in both experimental conditions.} 
Estimates are obtained from within-question linear regression models with homework-question fixed effects and cluster-robust standard errors clustered at the student level. Feedback provision is estimated using a linear probability model. For feedback length, submissions without feedback are assigned a value of zero. \revision{Feedback time, adjusted feedback time}, and feedback ratings are restricted to submissions with observed values and should therefore be interpreted descriptively rather than causally. \revision{Feedback stage timing data are missing for some submissions because some TAs initially used an earlier version of the extension from the HW0 pilot that did not record timing information; this issue was corrected once detected.}}
\label{tab:regression_results}
\begin{tabular}{lccccc}
\toprule
\textbf{Outcome} & \textbf{Estimate} & \textbf{SE} & \textbf{95\% CI} & $p$ & $N$ \\
\midrule
Feedback Provision 
& 10.81 
& 1.10
& [8.60, 13.00] 
& $<.001^{***}$ 
& 2{,}828 \\

Feedback Length 
& 39.79 
& 3.45
& [33.03, 46.55] 
& $<.001^{***}$ 
& 2{,}828 \\

\revision{Feedback Stage Time} 
& \revision{10.55} 
& \revision{4.19} 
& \revision{[2.35, 18.76]} 
& \revision{$.012^{*}$} 
& \revision{2{,}461} \\

\revision{Feedback Time} 
& \revision{25.35} 
& \revision{15.85} 
& \revision{[-5.72, 56.41]} 
& \revision{0.11} 
& \revision{664} \\

\revision{Adjusted} Feedback Time 
& 0.29 
& 0.35 
& [-0.39, 0.97] 
& 0.41 
& 664 \\

Feedback Rating 
& -0.01 
& 0.06 
& [-0.13, 0.12] 
& 0.88
& 468 \\
\bottomrule
\end{tabular}
\end{table}

\subsection{RQ1 -- Feedback Provision Behavior}
We report our regression results in Table~\ref{tab:regression_results} and provide summary statistics for outcomes observed in both experimental conditions in Appendix~\ref{app:summary_metrics} (see Tables~\ref{tab:shared_summary}-\ref{tab:shared_stratified_3}).

\xhdr{AI assistance increases feedback provision and feedback length.} 
We find that treatment assignment increased feedback provision by 10.81 percentage points (SE = 1.10, 95\% CI [8.60, 13.00], $p < 0.001$) and feedback length by 39.79 characters (SE = 3.45, 95\% CI [33.03, 46.55], $p < 0.001$). Because submissions without feedback are coded as zero characters, the feedback-length estimate captures the combined effect of increasing both feedback provision and the amount of feedback students receive. These patterns are consistent across homework assignments, \revision{with the treatment effect on feedback provision remaining positive and statistically significant for each homework assignment individually (8.21-12.60 percentage points, all $p < 0.001$)}.

\xhdr{\revision{AI assistance shifts rather than eliminates the labor of feedback provision.}}
\revision{We find that treatment assignment significantly increased time spent in the feedback stage across submissions (coeff. = 10.55 s, SE=4.19, 95\% CI [2.35, 18.76], $p=.012$). Among submissions that received feedback, however, differences in feedback time (coeff. = 25.35 s, SE = 15.85, 95\% CI [-5.72, 56.41], $p=0.11$) and adjusted feedback time (coeff. = 0.29 s/char, SE = 0.35, 95\% CI [-0.39, 0.97], $p = 0.41$) were not statistically significant. Thus, while AI assistance increases both the amount of feedback provided and its length, it does not reduce the measured time spent in the feedback stage. Although the increase in feedback stage time may partly reflect the greater likelihood of providing feedback, AI assistance can also introduce additional work, such as ``proof-read[ing]'' (T7) and ``changing the wording'' (T2) of the drafts. One interpretation is that AI assistance lowers the barrier to initiating feedback rather than eliminating the labor involved in providing it, which we further unpack  through TA usage patterns and interviews in Section~\ref{sec:rq2_results}.}

% ORIGINAL VERSION
% We find no statistically significant effect of treatment assignment on time spent per character of feedback (coeff. = 0.29 s/char, SE = 0.35, 95\% CI [-0.39, 0.97], $p = 0.41$). Although average time per character is higher in the treatment condition at the course level (Appendix~\ref{app:summary_metrics}), this difference is not statistically significant after accounting for homework-question fixed effects. Thus, while AI assistance increases both the amount of feedback provided and its length, we find no evidence that it reduces the measured time per unit of feedback. One interpretation is that AI assistance lowers the barrier to initiating feedback, but does not eliminate the work of reviewing, verifying, adapting, and deciding whether to send feedback. This interpretation is further supported by the usage patterns and interview findings reported in Section~\ref{sec:rq2_results}.

\begin{tcolorbox}[colback=gray!5,colframe=black,boxrule=0.5pt]
\xhdr{\textit{Takeaway.}} 
\textit{AI-assisted feedback drafts increase both feedback provision and feedback length, \revision{but we find no evidence that they reduce the measured time spent in the feedback stage}. \revision{One interpretation of this pattern} is  that AI assistance primarily supports feedback initiation, while TAs still expend effort reviewing, adapting, and deciding whether to send feedback.}\end{tcolorbox}

\subsection{RQ2 -- Usage Patterns}
\label{sec:rq2_results}

We report summary statistics for our treatment-only outcomes at the homework- and question-level in Table~\ref{tab:ta_experience} and by subgroup in Table~\ref{tab:ta_experience_stratified}, complemented by findings from our TA interviews. Throughout the interviews, the terms ``draft" and ``template" were used interchangeably, and some participant quotes therefore use the term ``template".

\begin{table*}[b]
\centering
\small
\caption{\textbf{Course- and homework-level summaries for treatment-only outcomes.} Feedback drafts are used in a considerable number of cases and exhibit moderate similarity with the final feedback, indicating that TAs adapt rather than copy the feedback drafts. Values are reported as proportions or means as appropriate; standard errors (SEs) are shown in parentheses. 
}
\label{tab:ta_experience}
\begin{tabular}{lcccc}
\toprule
\textbf{Level} & \textbf{Draft Usage \%} & \textbf{Draft Rating Avg} & \textbf{Similarity Score Avg} & \textbf{Grade Reconsideration \%} \\
\midrule
HW1    & 59.75 (3.90) & 3.38 (0.06) & 0.61 (0.03) & 5.05 (1.23) \\
HW2    & 78.41 (4.41) & 3.08 (0.05) & 0.66 (0.03) & 1.14 (0.57) \\
HW3    & 43.07 (4.25) & 2.84 (0.06) & 0.53 (0.03) & 6.28 (1.24) \\
HW4    & 33.78 (5.53) & 2.76 (0.05) & 0.62 (0.05) & 3.05 (0.91) \\
\midrule
\textbf{Course} & \textbf{54.15 (2.33)} & \textbf{3.00 (0.03)} & \textbf{0.60 (0.02)} & \textbf{3.90 (0.52)} \\
\bottomrule
\end{tabular}
\end{table*}

\xhdr{AI assistance supports verification and articulation.}
We find that feedback drafts occasionally prompted grade reconsideration, with TAs reporting that they would have assigned a lower grade after seeing the drafts in 3.90\% of submissions at the course level. Our interview data further help explain how the drafts supported both verification and feedback articulation. TAs frequently described difficulty in articulating feedback, even when they could identify issues in a submission. In these cases, the feedback drafts, whose personalization and specificity were much appreciated by some TAs (T1, T7, T9, T11), helped translate implicit judgments into clear feedback and made it easier for TAs to articulate and thus initiate feedback provision even when they already knew what they wanted to communicate (T1-T11):
\begin{quote}
    \textbf{T1}: \textit{``I really found it helpful because a lot of times, I wouldn't know how to give them the feedback the right way or put it in their own words, and the templates would do it in such a good way."}

    \vspace{0.2cm}
    \textbf{T4}: \textit{``I think in some cases, especially when a student got a question wrong, and I was trying to explain where they made a wrong step, the template was useful in pinpointing that out for me [...] it was very helpful to see how the template would word the feedback."}
\end{quote}

Beyond supporting articulation, TAs also described using feedback drafts as a verification mechanism that increased confidence in their decisions. Feedback drafts often helped identify errors that TAs initially missed, particularly in long or complex submissions (T1-T7, T11):

\begin{quote}
    \vspace{0.1cm}
    \textbf{T6}: \textit{``One of the things that was really cool about the template, and which I kinda enjoyed, it was something like an attention check. Because it's like you're required to do it, so I anyway have to click on. And so sometimes, if I get lazy with grading, and I may have a very long answer, even though overall it might look correct, I may have missed some minor points, and sometimes the template would catch that."}

    \vspace{0.2cm}
    \textbf{T7}: \textit{``That happened to me multiple times when the template was able to pick up an error that I missed, which was amazing, because then I actually went back and saw that it was incorrect."}
    \vspace{0.1cm}
\end{quote}

\xhdr{TAs adapt rather than copy feedback drafts.}
We find moderate lexical overlap between the feedback draft and the final feedback from TAs, with an average ROUGE-L F1 score of 0.60 at the course level. This suggests that TAs often use the feedback draft as a starting point, adapting it to better fit the specific context of a given submission rather than copying it verbatim. This idea of AI-assisted drafts serving as intermediate artifacts that guide feedback construction while preserving TA judgment was echoed by the TAs themselves. Rather than copying the drafts directly, TAs also described selectively using and modifying their content (T1, T5-T10): 
\begin{quote}
    \vspace{0.1cm}
    \textbf{T2}: \textit{``I kind of get started with a template and then fine-tune it as we go on."}

    \vspace{0.2cm}
    \textbf{T4}: \textit{``Even though I really cut down on the templates a lot, it was very helpful to see how the draft would word the feedback."}
\end{quote}
% This pattern suggests that while the feedback drafts help TAs get started, they do not eliminate the need for active engagement in the feedback process, suggesting once again that AI assistance reduces the initiation cost of feedback provision rather than the total effort required to give feedback.

\begin{table*}[t]
\centering
\small
    \caption{\textbf{Course-level summaries by subgroup for treatment-only outcomes.} \revision{TAs use the feedback drafts more on applied questions and rate them higher for submissions with mistakes}, with usage rates also varying by TA role. Values are reported as proportions or means as appropriate; standard errors (SEs) are shown in parentheses.}
\label{tab:ta_experience_stratified}
\begin{tabular}{llccc}
\toprule
\textbf{Category} & \textbf{Group} & \textbf{Draft Usage \%} & \textbf{Draft Rating Avg} &  \textbf{Similarity Score Avg} \\
\midrule
\multirow{2}{*}{\textbf{Mistake}}
& No Mistakes   & 57.78 (3.69) & 2.85 (0.03) & 0.61 (0.03) \\
& Has Mistakes  & 51.80 (3.00) & 3.51 (0.07)  & 0.59 (0.02)\\
\midrule
\multirow{2}{*}{\textbf{Question}}
& Applied    & 64.46 (3.30) & 3.06 (0.04) & 0.67 (0.03) \\
& Conceptual & 45.34 (3.17) & 2.94 (0.04) & 0.54 (0.02) \\
\midrule
\multirow{2}{*}{\textbf{Role}}
& GTA & 82.00 (3.15) & 2.79 (0.04) & 0.76 (0.03) \\
& UTA & 40.58 (2.80) & 3.25 (0.04) & 0.51 (0.02) \\
\bottomrule
\end{tabular}
\end{table*}

\xhdr{Feedback drafts vary in usefulness across grading contexts.}
We find that TAs' perceived usefulness of the feedback draft varies across grading contexts. For instance, the drafts receive higher usefulness ratings (3.51 vs. 2.85) on submissions with mistakes, suggesting that TAs find them more useful when feedback is clearly warranted. Similarly, the drafts are used more frequently (64.46\% vs. 45.34\%) and rated higher for usefulness (3.06 vs. 2.94) on applied questions than for conceptual ones, potentially reflecting the more open-ended nature of conceptual responses. We also observe differences across TA roles. GTAs use templates more frequently (82.00\%) but rate them lower for usefulness (2.79), while UTAs use them less often (40.58\%) but rate them higher for usefulness (3.25). 
These patterns suggest that perceptions of feedback draft usefulness may vary across grader experience levels and grading contexts:

\begin{quote}
    \vspace{0.1cm}
    \textbf{T8}: \textit{``If the student did the question correctly, I won't provide feedback. I think this is kind of reasonable. Since if you got credit, you don't look at the comments or feedback."}

    \vspace{0.2cm}
    \textbf{T11}: \textit{``Maybe it was good in one sense that it was a sanity check, like, oh yeah, I thought it was right, and the template thought it was right. But I think it was only really useful when [the submission] was wrong, and then we could align that we think it's wrong in the same way."}
\end{quote}

\revision{Taken together, these findings help contextualize the overall average draft usefulness rating of 3 (the midpoint of the five-point scale). While TAs found drafts useful in certain grading contexts, particularly when submissions contained mistakes, they provided less additional value in others.}

\begin{figure*}[t]
\centering
\includegraphics[width=0.875\textwidth]{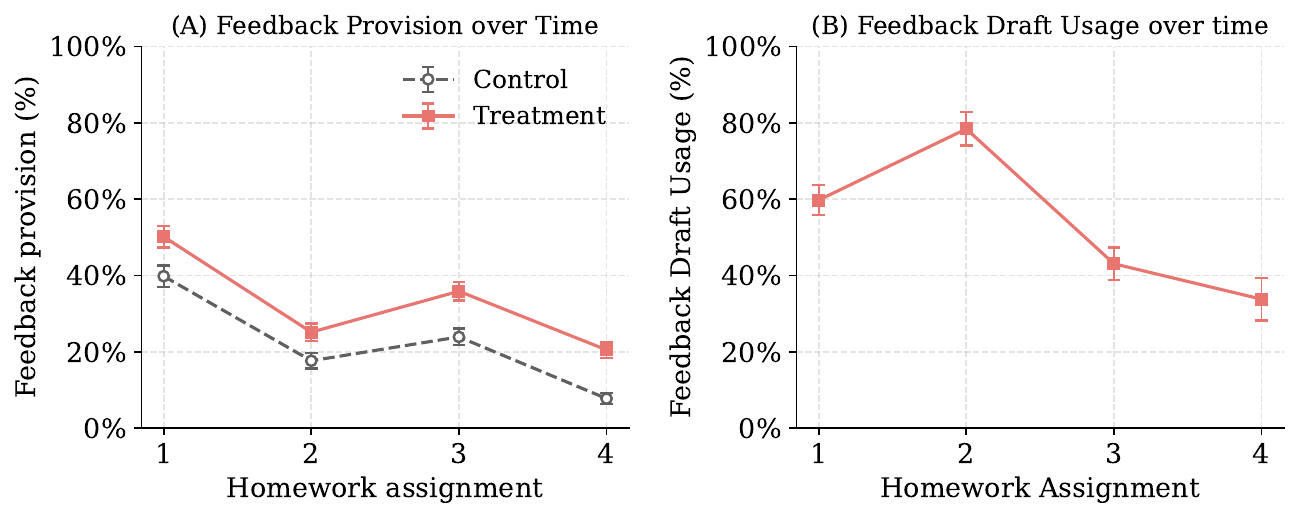}
\caption{\textbf{Temporal dynamics of AI-assisted feedback provision and feedback draft usage over the semester.} 
(A) AI assistance consistently increases feedback provision across all four homework assignments, though feedback provision declines over time in both conditions as the semester progresses. (B) Feedback draft usage is initially high but decreases over later assignments, suggesting that AI-assisted drafts may primarily support early-stage task initiation and articulation. Error bars denote standard errors.
}
\label{fig:temporal_decline}
\end{figure*}

\xhdr{Feedback draft usage and usefulness decline over time.}
At the course level, feedback drafts were used in 54.15\% of \revision{the final feedback provided by TAs to students} and received an average usefulness rating of 3.00. However, both usage and usefulness ratings decline over time, with usage decreasing from 59.75\%  to 33.78\%, and ratings decreasing from \revision{3.38} to 2.76. This mirrored the overall decline in feedback provision observed throughout the semester (see Figure~\ref{fig:temporal_decline}), \revision{even though the treatment effect on feedback provision remained positive and statistically significant for each homework assignment individually (8.21-12.60 percentage points, all $p < 0.001$).}

\revision{Our interviews suggest some explanations for this decline. First, AI-generated drafts introduced additional labor. For example, T7 mentioned that \textit{``it takes me longer to also proof-read the template to make sure that the issue it's identifying is correct,''} while T2 sometimes found grading faster without the draft because of the editing it required. As the semester progressed, competing priorities could also make this additional effort less worthwhile. T4, for example, mentioned that toward the end of the semester, when they \textit{``just [had] to get this done,''} they would stop reading the templates and focus solely on grading. The scale of the course further complicated sustained feedback provision, with multiple TAs mentioning how the large course size made it difficult to consistently give feedback to everyone (T7, T10). In response to the added labor and competing priorities, T6 suggested that sustaining feedback provision may require additional incentives:}

\begin{quote}
    \vspace{0.1cm}
    \textbf{T6}: \textit{``There needs to be some incentive [...] Maybe like using it allows you grade only 50\% of questions and the other 50\% are automatically graded. That's an incentive, a big incentive. I don't know if that's the right incentive, or if that's something which you would want to do, but maybe there are some specific questions where that's fine."}
\end{quote}

\revision{As such, even though AI assistance may make it easier to initiate feedback provision, it does not necessarily provide enough support to sustain that participation over time. }
% \revision{Although its treatment advantage persisted, the additional labor of reviewing and editing drafts, competing priorities, and the scale of the course may have limited its ability to prevent its overall decline.}

\begin{tcolorbox}[colback=gray!5,colframe=black,boxrule=0.5pt]
\xhdr{\textit{Takeaway}}. 
\textit{\revision{AI-assisted feedback drafts help TAs initiate feedback by providing editable scaffolds, but do not eliminate the labor of reviewing, verifying, and editing responses or necessarily reduce the ongoing effort required to sustain feedback provision, particularly amid competing priorities and the demands of providing feedback at scale.}}\end{tcolorbox}

% ORIGINAL VERSION
% \begin{tcolorbox}[colback=gray!5,colframe=black,boxrule=0.5pt]
% \xhdr{\textit{Takeaway}}. 
% \textit{AI-assisted feedback drafts function as editable scaffolds: they help TAs initiate feedback by articulating, checking, and refining responses without replacing TA judgment. TAs often adapt these drafts rather than copy them verbatim, and usage declines over time, suggesting that AI assistance lowers the initial barrier to feedback provision but does not eliminate the ongoing effort required to sustain it.}\end{tcolorbox}

\subsection{RQ3 -- Downstream Experiences}
We report our regression results for student usefulness ratings in Table~\ref{tab:regression_results} and provide their corresponding summary statistics across both conditions in Appendix~\ref{app:summary_metrics} (see Tables~\ref{tab:shared_summary} and~\ref{tab:shared_stratified_3}).
Because treatment assignment was randomized, these comparisons are informative about whether feedback produced under the AI-assisted workflow was rated less useful than feedback produced under the control workflow. However, ratings are only observed for feedback that students received and evaluated. As a result, the analysis should be interpreted as applying to rated feedback instances, not as an unconditional estimate of the effect of AI assistance on all student experiences.

\xhdr{AI assistance does not reduce student perceived usefulness of rated feedback.}
Among submissions with observed student usefulness ratings, perceived usefulness was nearly identical across conditions (coeff. = -0.01, SE = 0.06, 95\% CI [-0.13, 0.12], $p = 0.88$). Course-level averages were also similar: 3.47 in the control and 3.48 in the treatment, with similar patterns across homework assignments. \revision{As such, among submissions with observed student ratings}, we find no evidence that feedback produced with AI assistance was perceived as less useful by students, even as the treatment increased its provision and length.

\xhdr{\revision{AI assistance maintains expert perceived quality of provided feedback.}}
\revision{Experts generally preferred the final provided feedback from the treatment group, with the primary reason for this being attributed to specificity. Across the 105 post-calibration pairs, experts preferred treatment feedback in 61.9\% of cases (65/105), control feedback in 3.8\% (4/105), and expressed no preference in 34.3\% (36/105). A similar pattern was observed across all three experts, with the three experts each having majority preference for treatment feedback (48.6\%, 68.6\%, 68.6\%), followed by no preference (40.0\%, 31.4\%, 31.4\%), and followed by control (11.4\%, 0.0\%, 0.0\%).}

\revision{Across the 65 pairs in which treatment feedback was preferred, most preferences were mild (84.6\%), with a few cases receiving a strong preference (15.4\%). Specificity was the most frequently attributed factor for preferring treatment feedback (90.8\%), followed by pedagogical appropriateness (15.4\%), correctness (12.3\%), and actionability (10.8\%). All four preferences for control feedback were mild and attributed to specificity. No additional reasons for preference were reported, and experts identified no issues or concerns with either treatment or control feedback.}

\xhdr{Students value increased access to specific and actionable feedback.} 
Student interviews help explain this similarity in perceived usefulness. Rather than substantially changing the perceived quality of individual feedback instances, AI assistance appears to increase the \textit{availability of feedback} while \revision{ still resulting in finalized, TA-reviewed feedback with} the characteristics students valued, such as feedback that was specific, actionable, and targeted to their mistakes, particularly when it clearly identified where they went wrong and how to improve (S2, S3, S6-S8). Students also appreciated receiving more detailed and personalized feedback than in other courses, especially when the feedback went beyond generic rubric categories (S2-S5, S8, S9):

\begin{quote}
    \vspace{0.1cm}
    \textbf{S6}: \textit{``In general, I found that most responses to questions I got wrong were pretty helpful. Especially ones that pointed out where the grader believed that I went wrong in the understanding process."}
    
    \vspace{0.2cm}
    \textbf{S9}: \textit{``Out of all the courses I've taken, most of them don't provide any feedback at all. This [course] is the one that provides the most, and I think that's very helpful."}
\end{quote}

\xhdr{Students reported difficulty distinguishing AI-assisted from non-assisted feedback and preferred human oversight.} In interviews, students generally reported that they could not tell which feedback was AI-assisted and did not spontaneously express reduced trust in the feedback they received (S2, S3, S5, S6--S9). At the same time, students expressed a preference for human-in-the-loop feedback over fully automated feedback, emphasizing the importance of human oversight and expressing hesitation toward feedback generated and delivered entirely by AI systems (S1--S9):

\begin{quote}
\textbf{S1}: \textit{``I wouldn't be comfortable with feedback purely by an LLM, because we just know that they make mistakes, and sometimes they hallucinate, so that's the one I would be the least comfortable with."}

\vspace{0.2cm}
\textbf{S2}: \textit{``I don't know to what extent it was assisted by LLM, but no, I could never tell that it was an LLM or template or anything."}
\end{quote}

\begin{tcolorbox}[colback=gray!5,colframe=black,boxrule=0.5pt]
\xhdr{\textit{Takeaway}}. 
\textit{AI assistance increased access to feedback \revision{while maintaining both student perceived usefulness and expert perceived quality among the feedback instances for which these outcomes were observed}. Student interviews suggest that students valued feedback that was specific, actionable, and personalized, while also preferring AI-assisted feedback workflows that preserved human oversight.}\end{tcolorbox}

\section{Discussion}
\revision{Our study shows that AI assistance did not simply make feedback provision easier or faster. It instead changed participation in a communicative practice between TAs and students without evidence of reduced student-rated usefulness or expert-perceived quality among the feedback provided. While prior work on AI-mediated communication has examined how AI changes the content and interpersonal nature of communication~\cite{Mieczkowski21AIMC,Zhang22AITranslationsMeetings,Sun24AIPeerReview}, our findings highlight how AI assistance can also change if discretionary communication happens at all. TAs provided feedback that would have otherwise gone unwritten, while having to still bear the work of engaging in and reviewing that communication,  highlighting how the benefits and labor of AI-assisted communication can be distributed across different actors. The persistence of this participation gain across the semester along with the overall decline in provision also shows that helping people initiate discretionary work is distinct from making that work sustainable, which may require additional forms of support. Here, we further discuss how and why AI support changes the dynamics of discretionary work as well as the corresponding design implications.}

% ORIGINAL VERSION
% Our study shows that AI assistance can meaningfully change participation in discretionary work even when it does not substantially reduce measured effort per unit of work. In a semester-long randomized field experiment, AI-assisted feedback drafts increased both the provision of feedback and its length. Yet we found no evidence that they reduced measured time per character, and usage declined over the semester. Interview data help explain this pattern: TAs used drafts to articulate feedback, check their judgments, and structure responses, but still needed to review, adapt, and decide whether to send the feedback. Students, in turn, rated feedback produced under the AI-assisted workflow as similarly useful to feedback in the control condition and expressed a preference for maintaining human oversight. Together, these findings suggest that AI-assisted systems may be especially valuable for discretionary but beneficial work, as they make it more likely that people begin and engage with work that might otherwise be skipped.

\subsection{\revision{AI Assistance Can Increase Participation Without Reducing Labor}}
Much prior work on AI-assisted workflows evaluates tasks that users are already expected to complete, emphasizing productivity, quality, accuracy, or efficiency~\cite{Noy23,Peng23,Cui25,Zhang24}. Our findings highlight a different design problem: in discretionary work, the central question is often whether the task happens at all. This distinction connects to longstanding CSCW concerns about invisible and voluntary work, where collaborative systems depend on labor that is valuable to collective outcomes but weakly rewarded or easy to overlook~\cite{grudin1988cscw, star1999layers,schmidt1992taking,Kow18,Meluso25}. 

\revision{In Course X, feedback provision was optional and historically rare. In this discretionary setting, we find that AI-assisted drafts increased whether TAs provided feedback and how much feedback students received, with expert evaluations suggesting that the final feedback produced under the AI-assisted workflow maintained quality and was often preferred. At the same time, AI assistance did not reduce the labor dedicated to giving feedback: treatment increased total time spent in the feedback stage as TAs continued to review, verify, and edit the drafts. As such, AI assistance can produce participation gains for optional tasks without reducing the labor required for it. Our qualitative findings suggest one mechanism through which this may occur. The AI-assisted drafts provided an editable starting point that often helped TAs articulate thoughts they already had in mind and also verify their own judgments. Rather than eliminating the feedback work, the AI assistance may have lowered the initial barrier to providing feedback, increasing participation without producing labor savings.}

% ORIGINAL VERSION
% In Course X, feedback provision was optional and historically rare. AI assistance increased both the likelihood that TAs provided feedback and the amount of feedback students received, even though it did not significantly reduce measured time per character. This suggests that AI assistance can produce \textit{participation gains} without necessarily producing conventional efficiency gains. In our setting, drafts appeared to lower initiation barriers for providing feedback by giving TAs a concrete starting point, but TAs still had to check, adapt, and decide whether to send the feedback. If evaluations focus only on speed or quality conditional on task completion, they may miss this form of impact: AI systems can matter by making optional but beneficial work more likely. Similar dynamics may arise in peer review, code review, documentation, mentoring, advising, moderation, and other forms of collaborative labor that are valuable but easy to defer.

\subsection{\revision{From Initiating to Sustaining Discretionary Participation}}
Although AI assistance increased feedback provision \revision{and maintained the treatment effect across all four homeworks}, absolute feedback provision and draft usage still declined over the semester. \revision{In other words, AI assistance continued to increase the probability of feedback provision relative to no assistance, but neither condition was left unaffected by the overall labor required by giving feedback as well as semester-level pressures. During interviews, TAs described how, although AI-generated drafts provided helpful starting points, they still required review and verification. As competing responsibilities built up over the semester, TAs found it difficult to prioritize this additional labor, particularly given the lack of incentives for providing feedback.}

This suggests that lowering the immediate friction of discretionary work may help users get started but may not be sufficient to sustain participation when the work remains effortful, weakly incentivized, or institutionally underrecognized. This pattern echoes longstanding CSCW concerns about invisible and voluntary work: valuable collaborative labor often remains difficult to measure or reward~\cite{star1999layers}, and systems can struggle when the people who bear the costs of participation are not the same people who receive its benefits~\cite{grudin1988cscw}. 
In our setting, AI drafts helped TAs begin feedback, but they did not remove the need to review, verify, and refine drafts amid broader semester workloads. Sustaining discretionary work may therefore require more than AI assistance alone. Future systems may need to pair AI-assisted scaffolds with social and organizational supports, such as clearer expectations, recognition, incentives, or workload allocation, that make valuable but underrecognized work easier to sustain over time~\cite{kraut2012building,Kow18,Meluso25}.

\subsection{\revision{Implications \& Future Directions for AI Assistance in Discretionary Work}}

Our findings suggest several implications for the design of AI-assisted workflows in discretionary settings while also raising questions about whether participation gains from AI assistance transfer to other forms of discretionary work.

\xhdr{Design for participation, not only for efficiency}. We find that AI assistance can increase discretionary participation without necessarily reducing the labor required by it. With this in mind, designers should consider whether AI systems encourage valuable contributions that might otherwise go unmade, rather than focusing only on task efficiency. Future work should also explore whether similar participation gains apply in other contexts, such as courses with different pedagogical structures and other forms of discretionary but socially valuable work. 
    
\xhdr{Support the remaining human labor}.  Although AI-assisted drafts increased feedback provision, TAs still remained responsible for reviewing, verifying, and refining them. As such, designers should support this remaining labor through features that facilitate verification, editing, and selective use of assistance. Future work could also compare alternative forms of AI assistance and explore whether they impact the involved work and labor differently.
    
\xhdr{Design for sustained participation}. While AI assistance consistently increased feedback provision, overall participation declined over time. This suggests that reducing barriers to initiating discretionary work may be insufficient for sustaining it. Accordingly, an important direction for future work is to consider how technical assistance can complement organizational support, including appropriate incentives and recognition, rather than relying on AI assistance alone.

\section{Limitations}
\label{sec:limitations}

Our findings should be interpreted in light of several limitations related to measurement, intervention design, and generalizability.

\xhdr{Measurement.} 
Our interpretation that AI assistance may support task initiation rather than reducing total effort is based on indirect evidence. This interpretation is informed by behavioral patterns (e.g., increased feedback provision and length without reducing the time spent on feedback) and TA interviews. However, we do not directly measure initiation costs, cognitive effort, or the work required to verify and adapt feedback drafts \revision{and thus cannot establish reduced initiation barriers as the mechanism underlying our effects.} Some outcomes also rely on self-reports, including TA usefulness ratings of feedback drafts and student usefulness ratings of the feedback they received, which may be subject to reporting biases or interpretive variation. Finally, student usefulness ratings are observed only when students received feedback and completed the optional post-homework survey. \revision{Because treatment affects whether feedback is provided and survey participation is voluntary, these ratings may be subject to selection bias and not representative of the overall student experience.} These ratings should therefore be interpreted as evidence about rated feedback instances rather than as an unconditional measure of all student experiences.

\xhdr{Intervention Design.} 
Our intervention intentionally introduced AI assistance only after grading was completed. This design helped preserve independent TA judgment and reduce over-reliance, but it also means our findings reflect a particular form of AI assistance: feedback drafts presented after evaluation, for optional use in providing feedback. Other designs that integrate AI earlier in the pipeline, provide stronger recommendations, or automate larger portions of the workflow may thus produce different effects. \revision{Moreover, since the message displayed by our extension differs between conditions, our design does not fully separate the effect of receiving an AI-assisted feedback draft from the potential nudging effects of the different messages, which may have independently influenced TAs' decisions to provide feedback.} 

\revision{More broadly, our study compares AI-assisted feedback drafts against no drafts, rather than alternative forms of assistance such as generic templates, reminders, or suggested starting points. As such, we cannot isolate whether the increase in feedback provision is specific to AI personalization or could also be achieved through these simpler forms of assistance.}
Lastly, because treatment and control conditions were assigned at the question level and interspersed throughout grading, some treatment effects may have spilled over into the control condition. Several TAs noted that they would occasionally ``save'' drafts they found useful and later adapt them for other submissions, including potentially control submissions, particularly when students made similar mistakes. As a result, our estimated treatment effects may be conservative, and the true effects of AI-assisted feedback drafts may in fact be larger.

\xhdr{Generalizability.} 
The study was conducted in a single upper-level machine learning course at a private university, using a single model configuration, prompt, and grading platform integration. This setting provides ecological validity because the intervention was embedded in a real course workflow, but results may differ in other domains, institutions, or courses where feedback expectations, grading practices, or AI norms differ. The study also spans a single semester. Although this duration allowed us to observe a temporal decline in draft usage and feedback provision, longer-term studies are needed to understand whether AI-assisted feedback becomes normalized, abandoned, or reshaped as instructors and TAs gain experience with the system.

\section{Conclusion}
We study AI assistance in a higher-education setting where personalized feedback is discretionary and often underprovided. In a randomized field experiment, AI-assisted feedback drafts increased both feedback provision and length, while student-rated usefulness remained comparable across conditions and \revision{experts generally preferred treatment feedback}. Rather than replacing TA judgment, the drafts functioned as editable scaffolds that TAs reviewed, adapted, and refined. However, \revision{although the treatment effect on feedback provision persisted throughout the semester}, overall feedback provision and draft usage declined over time. \revision{Together with our timing and interview findings}, this suggests that AI assistance can lower barriers to initiating discretionary work without eliminating the labor required to sustain it, \revision{and may introduce additional work through reviewing and refining AI-generated drafts}. Taken together, our findings shift the focus from AI-assisted productivity gains to AI-assisted participation gains. For discretionary but socially beneficial practices, such as feedback, mentoring, documentation, and other forms of often-invisible collaborative labor, AI systems may matter not only because they make work faster, but because they make valuable work more likely to happen. At the same time, sustaining such work likely requires more than AI assistance alone; it also requires incentives, recognition, and organizational support.

\section*{Acknowledgements}
This work was made possible through the funding provided by the Princeton University's Humanities Council and the Keller Center for Innovation in Engineering Education.

\section*{Generative AI Disclosure}
Generative AI tools were used to enhance the search for related works and refine the writing and formatting of this manuscript. Specifically, Claude, ChatGPT, and Elicit were used to find relevant research papers for both the related works and discussion sections (alongside non-Generative AI tools, like Google Scholar). After the Discussion had been written, ChatGPT was used to refine and streamline the content, and then manually edited again by the authors. Claude was also used for specific formatting tasks, such as generating table formats. Where generative AI has been used for editing, the authors certify that they have read, adapted, and corrected the text as necessary, and stand behind the resulting work.

\newpage
\bibliographystyle{ACM-Reference-Format}
\bibliography{references}

\newpage
\appendix
\section{Sample Problems from Course X}
\label{app:sample_problems}
We provide one sample problem for each of the question types provided in the Course X below.

\subsection{Applied Question Sample Problem}

\begin{tcolorbox}[colback=gray!5,colframe=black,boxrule=0.5pt]
Consider the following training dataset:
\begin{center}
    \begin{tabular}{l|l}
        $x$ & $y$ \\ 
        \hline
        1 & 4 \\
        3 & -4 \\
        5 & 6 \\
        7 & 2 \\
    \end{tabular}
\end{center}

\noindent We would like to fit a linear regression model $\hat{y} = w_0 + w_1x$ to this dataset using gradient descent and $\ell_2$ regularization, with $\lambda = 0.2$ and $\eta = 0.1$. We use mean squared error for our loss function and initialize $\vec{w} = (w_0, w_1)$ with $(0, 0)$.

\begin{enumerate}
    \item Perform the first two gradient updates to $\vec{w}$. Round to three decimal places.
    \item What would happen if we set $\eta$ to be much smaller? What about much larger? 
\end{enumerate}
\end{tcolorbox}

\subsection{Conceptual Question Sample Problem}
\begin{tcolorbox}[colback=gray!5,colframe=black,boxrule=0.5pt]
If a distribution $\mathcal{D}$ over a finite set $\{ 1,2,\dots,|V|\}$ assigns probability $p_i$ to element (number) $i$, then the entropy of the distribution is defined to be $\sum_i p_i \log (1/p_i)$. 
For the rest of the question, let $V$ denote the number of words in the English dictionary and $p_i$ denote the unigram probability of the $i^{th}$ word in the dictionary. 
Let the entropy $\sum_i p_i \log (1/p_i)$ represent the \textit{word entropy of English}.

\vspace{0.2cm}
By definition, if we look at a corpus of length $T$, where $T\to \infty$, then we expect to find the word $i$ occurring close to $p_i \cdot T$ times (formally, this follows from the central limit theorem, but you can assume it to be true). 

\vspace{0.2cm}
Using this fact, show that the logarithm of the perplexity of a unigram language model on a large held-out corpus $\mathcal{C}$ is the \textit{word entropy of English}.
\end{tcolorbox}

\newpage
\section{Prompt for Feedback Provision}
\label{app:feedback_prompt}

We provide our prompt for generating AI-assisted feedback drafts below.

\begin{tcolorbox}[
  colback=gray!5,
  colframe=black,
  boxrule=0.5pt
]
\ttfamily
You are a companion to a teaching assistant (TA) in an undergraduate-level computer science course on machine learning.
Your task is to help the TA generate actionable, constructive feedback on a student's submission to a problem.

\vspace{0.2cm}
For a \textbf{submission with mistakes}, feedback should identify the errors clearly and provide guidance on how the student can address them.
For a \textbf{submission without mistakes}, the feedback should provide concise positive reinforcement, highlighting what was done particularly well (if applicable) and pointing out minor errors/inconsistencies that didn't get caught by the grading rubrics (if applicable).

\vspace{0.2cm}
Use a feedback strategy best suited to the submission. These strategies could include (but are not limited to):
\begin{itemize}
    \item Hints/explanations on terminology
    \item Examples illustrating the concept
    \item Hints on conceptual context or attributes
    \item Error identification and correction guidance
    \item Task-specific strategies or processing steps
    \item Guiding questions or worked-out examples
\end{itemize}

As part of this task, you are provided with:
\begin{itemize}
    \item Problem description
    \item Instructor's solution
    \item Instructor's grading rubric
    \item Student submission
\end{itemize}

Note that multiple correct approaches may exist for open-ended questions and that small typos are not considered as mistakes.  Evaluate based on alignment with the rubric and conceptual soundness, not strict similarity to the instructor’s solution.

\vspace{0.2cm}
Please provide concise feedback. Aim for \textbf{2--3 concise sentences} for a submission with mistakes or \textbf{1 concise sentence} for a submission without mistakes. Maintain a constructive and supportive tone and use the pronoun ``you" in your response to address the student.
When possible, try to embed the Unicode for math characters directly. Otherwise, use LaTeX formatting.

\vspace{0.2cm}
\#\#\# Problem Description: \{problem\}

\#\#\# Instructor's Solution: \{solution\}

\#\#\# Instructor's Grading Rubric: \{rubric\}

\#\#\# Student Submission: \{submission\}
\end{tcolorbox}

\newpage
\section{Sample Feedback Drafts}
\label{app:feedback_drafts}

We provided two example AI-assisted feedback drafts for the question given below. The first feedback, which was provided on a submission that had no mistakes, received a usefulness rating of 2/5, while the second feedback, which was provided on a submission that had mistakes, received a usefulness rating of 4/5. This was consistent with our observation that TAs found the feedback drafts more useful on submissions that had mistakes over those that did not.

\xhdr{Problem Statement}
\begin{tcolorbox}[colback=gray!5,colframe=black,boxrule=0.5pt]
One way of formally measuring the importance of a variable $x_i$ in computing a function $\ell$ is to look at the partial derivative $\frac{\partial \ell}{\partial x_i}$. Assume $\ell$ is the least squares loss function with respect to one test data point $(\vec{x}, y)$, and give an expression for $\frac{\partial \ell}{\partial x_i}$. Briefly explain why the weight $w_i$ can also be interpreted as the importance of $x_i$ in the model's decision.
\end{tcolorbox}

\xhdr{Sample Feedback 1}
Feedback on a submission to the above problem that did not have any mistakes.
\begin{tcolorbox}[colback=gray!5,colframe=black,boxrule=0.5pt]
Great work deriving the partial derivative and clearly showing how $w_{i}$ captures the sensitivity of the model’s score to $x_{i}$. You could further strengthen your answer by noting that the residual term is common across all features, making importance strictly proportional to $w_{i}$.
\end{tcolorbox}

\xhdr{Sample Feedback 2}
Feedback on a submission to the above problem that did have mistakes.
\begin{tcolorbox}[colback=gray!5,colframe=black,boxrule=0.5pt]
You correctly applied the chain rule to get $\partial l/\partial x_i = -2 w_i (y - \hat{y})$, but your explanation would be clearer if you pointed out that the $(y-\hat{y})$ term is the same for every coordinate, so the only thing that changes per $x_i$ is the factor $w_i$. Try dropping the $x_i<<1$ example and explicitly state that because the common term cancels out when comparing coordinates, $w_i$ alone governs the importance of $x_i$.
\end{tcolorbox}

\newpage 
\section{Feedback Quality Evaluation Instrument}
\label{app:feedback_quality_eval}
We provide our evaluation instrument for assessing feedback quality in matched control–treatment pairs below.

\medskip
\noindent\textbf{Pair X — Homework Y, Question Z}

\medskip
\noindent\textbf{Question Statement:} \{\{QUESTION\}\}

\medskip
\noindent\textbf{Student A}
\begin{quote}
\textbf{Submission:} \{\{ANSWER A\}\}

\textbf{Feedback:} \{\{FEEDBACK A\}\}
\end{quote}

\noindent\textbf{Student B}
\begin{quote}
\textbf{Submission:} \{\{ANSWER B\}\}

\textbf{Feedback:} \{\{FEEDBACK B\}\}
\end{quote}

\medskip
\noindent\textbf{Q1. Feedback Preference.} Considering each student's response, which piece of feedback do you have a preference for?
\begin{itemize}
    \item Strong preference for A
    \item Mild preference for A
    \item About the same
    \item Mild preference for B
    \item Strong preference for B
\end{itemize}

\noindent\textbf{Q2. Reasons for Preference.}
\textit{(Displayed only if a preference was indicated.)} What factors contributed to your preference of a piece of feedback over another? Select all that apply.
\begin{itemize}
    \item Correctness
    \item Specificity
    \item Actionability
    \item Pedagogical appropriateness
    \item Other (please specify)
\end{itemize}

\noindent\textbf{Q3. Issues \& Concerns.} Please note / comment on any issues or anything related to this pair as needed, such as whether there was answer leakage, hallucinations, or other concerns.

\newpage
\section{Interview Questions}
\label{app:interview_questions}
We provide the questions for our TA and students interviews below.
\subsection{Questions for TA Interviews}
\begin{itemize}
    \small
    \item How would you describe your overall experience with the AI-assisted feedback drafts?
    \item Did seeing the feedback drafts encourage you to provide feedback (even if you didn’t use them)?
    \item Would you use the feedback drafts unchanged or would you make edits? If so, were the edits minor or major? What types of edits would you make (e.g., cut-down, re-order, etc.)? 
    \item Did it take you longer to give feedback when you had feedback drafts or when you didn’t?
    \item When working with the feedback drafts, what would the majority of time be spent on?
    \item If you had the choice, would you choose to receive feedback drafts on all submissions or not receive them on any of the submissions? Why or why not?
    \item What else can we do to improve your feedback provision workflow and / or experience? 
    \item What did you like about feedback drafts? What didn’t you like about them?
    \item When did you find the feedback drafts useful? Not useful?
    \item Did you notice any recurring issues in the feedback drafts? 
    \item Did you notice any trends in the feedback drafts?
    \item How would you improve the quality of the feedback drafts?
\end{itemize}

\subsection{Questions for Student Interviews}
\begin{itemize}
    \small
    \item How did you find the feedback you received on homework assignments?
    \item Could you tell when a feedback was generated purely by a TA versus with the help of an LLM? If so, what were the major differences?
    \item Was there ever a time when you would have liked (more) feedback but didn’t receive it? If so, when? 
    \item Was there ever a time when you would have liked no (or less) feedback but did receive it? If so, when?
    \item How did you feel about receiving feedback on questions where you made a mistake? On questions where you didn’t make a mistake?
    \item How do you feel about receiving feedback that is (1) purely written by a GTA, (2) purely written by a UTA; (3) written by a GTA with the help of an LLM; (4) written by a UTA with the help of an LLM; or (5) purely written by an LLM? Would you be comfortable with all of these options? If not, why?
    \item What makes feedback useful to you in general? On questions where you made a mistake? On questions where you didn’t make a mistake?
    \item From all the courses you’ve taken, has there been a course that did particularly well with providing feedback on assignments? If so, what factors contributed to that experience?
\end{itemize}

\newpage
\section{\revision{Sample Sizes for Descriptive Subgroups}}
\label{app:subgroup_sample_size}

We provide the sample sizes for our descriptive subgroup categories in Table~\ref{tab:subgroup_sizes}. These categories include whether a submission has mistakes, whether a submission is for an applied or conceptual question, and whether a submission is graded by a GTA or a UTA.

\begin{table}[h]
\centering
\caption{Sample sizes across the descriptive subgroup categories across all submissions, in the treatment condition, and in the control condition.}
\label{tab:subgroup_sizes}
\begin{tabular}{llrrr}
\toprule
Category & Group & $N$ (Total) & $N$ (Treatment) & $N$ (Control) \\
\midrule
\multirow{2}{*}{Mistake}
    & No Mistakes & 2,194 & 1078 & 1116\\
    & Has Mistakes & 634  & 332 & 302\\
\midrule
\multirow{2}{*}{Question}
    & Applied & 1,453 & 730 & 723\\
    & Conceptual & 1,375 & 680 & 695\\
\midrule
\multirow{2}{*}{Role}
    & GTA & 1,522 & 755 & 767\\
    & UTA & 1,306 & 655 & 651 \\
\bottomrule
\end{tabular}
\end{table}

\newpage
\section{Summary Metrics for Outcomes Observed in Both Experimental Conditions}
\label{app:summary_metrics}

We provide summary metrics for feedback provision, feedback length, \revision{feedback stage time, feedback time, adjusted feedback time}, feedback rating at the course-level, homework-level, and by subgroup in Tables~\ref{tab:shared_summary},~\ref{tab:shared_stratified_1},\revision{~\ref{tab:shared_stratified_2}}, and~\ref{tab:shared_stratified_3}.

\begin{table}[h]
\centering
\small
\caption{\textbf{Course- and homework-level summaries for outcomes observed in both experimental conditions.} Across all homework assignments and at the course-level, AI assistance increases feedback provision and length while preserving the student-perceived usefulness of feedback. However, \revision{feedback stage time and feedback time are both longer in the treatment while adjusted feedback time remains similar across conditions.} Values are reported as proportions or means as appropriate; standard errors (SEs) are shown in parentheses.}
\label{tab:shared_summary}
\begin{tabular}{l l l l}
\toprule
\textbf{Outcome} & \textbf{Level} & \textbf{Control} & \textbf{Treatment} \\
\midrule
\multirow{5}{*}{\textbf{Feedback Provision \%}} 
 & HW1 & 39.81\% (2.76) & 50.16\% (2.81) \\
 & HW2 & 17.65\% (2.02) & 25.14\% (2.32) \\
 & HW3 & 23.90\% (2.18) & 35.86\% (2.46) \\
 & HW4 & 7.73\% (1.41) & 20.50\% (2.13) \\
 & \textbf{Course} & \textbf{21.72\% (1.10)} & \textbf{32.48\% (1.25)} \\
\hline
\multirow{5}{*}{\textbf{Feedback Length Avg (char)}} 
 & HW1 & 53.30 (5.91) & 123.62 (8.56) \\
 & HW2 & 8.83  (1.67) & 44.68 (4.80) \\
 & HW3 & 34.40 (4.56) & 74.70 (6.86) \\
 & HW4 & 6.78 (1.58) & 23.12 (3.96) \\
 & \textbf{Course} & \textbf{25.10 (1.95)} & \textbf{65.04 (3.25)} \\
\hline
\multirow{5}{*}{\textbf{\revision{Feedback Stage Time Avg (s)}}} 
 & \revision{HW1} & \revision{8.56 (3.08)} & \revision{14.20 (1.57)} \\
 & \revision{HW2} & \revision{2.91 (0.43)} & \revision{8.57 (0.97)} \\
 & \revision{HW3} & \revision{2.48 (0.25)} & \revision{27.02 (15.73)} \\
 & \revision{HW4} & \revision{1.99 (0.41)} & \revision{7.74 (1.93)} \\
 & \textbf{\revision{Course}} & \textbf{\revision{3.83 (0.70)}} & \textbf{\revision{14.36 (4.08)}} \\
\hline
\multirow{5}{*}{\textbf{\revision{Feedback Time Avg (s)}}} 
 & \revision{HW1} & \revision{12.90 (4.23)} & \revision{22.75 (2.97)} \\
 & \revision{HW2} & \revision{7.80 (1.93)} & \revision{17.43 (2.97)} \\
 & \revision{HW3} & \revision{4.96 (0.72)} & \revision{59.30 (39.11)} \\
 & \revision{HW4} & \revision{4.94 (2.45)} & \revision{12.99 (2.75)} \\
 & \textbf{\revision{Course}} & \textbf{\revision{8.34 (1.55)}} & \textbf{\revision{32.29 (12.76)}} \\
\hline
\multirow{5}{*}{\textbf{\revision{Adjusted} Feedback Time Avg (s/char)}} 
 & HW1 & 0.36 (0.06) & 0.18 (0.02) \\
 & HW2 & 0.42 (0.17) & 0.19 (0.04) \\
 & HW3 & 0.11 (0.02) & 1.02 (0.87) \\
 & HW4 & 0.09 (0.04) & 0.15 (0.04) \\
 & \textbf{Course} & \textbf{0.27 (0.05)} & \textbf{0.45 (0.28)} \\
 \hline
 \multirow{5}{*}{\textbf{Feedback Rating Avg}} 
 & HW1 & 3.60 (0.07) & 3.56 (0.08) \\
 & HW2 & 3.49 (0.13) & 3.46 (0.14) \\
 & HW3 & 3.29 (0.12) & 3.42 (0.13) \\
 & HW4 & 3.19 (0.15) & 3.38 (0.10) \\
 & \textbf{Course} & \textbf{3.47 (0.05)} & \textbf{3.48 (0.05)} \\
\hline
\end{tabular}
\end{table}

\begin{table}[h]
\centering
\small
\caption{\textbf{Course-level summaries by subgroup for feedback provision and length.} Submissions in the treatment show higher feedback provision and longer feedback across subgroups. Values are proportions or means as appropriate; standard errors (SEs) are shown in parentheses.}
\label{tab:shared_stratified_1}
\begin{tabular}{llcccc}
\toprule
& & \multicolumn{2}{c}{\textbf{Feedback Provision \%}} 
  & \multicolumn{2}{c}{\textbf{Feedback Length Avg (char)}} \\
\cmidrule(lr){3-4} \cmidrule(lr){5-6}
\textbf{Category} & \textbf{Group} 
& \textbf{Control} & \textbf{Treatment} 
& \textbf{Control} & \textbf{Treatment} \\
\midrule
\multirow{2}{*}{\textbf{Mistakes}}
& No Mistakes & 10.04 (0.90) & 16.70 (1.14) & 2.74 (0.44) & 19.87 (1.87) \\
& Mistakes    & 64.90 (2.75) & 83.73 (2.03) & 107.71 (7.29) & 211.72 (8.29) \\
\midrule
\multirow{2}{*}{\textbf{Question}}
& Applied    & 17.57 (1.42) & 28.90 (1.68) & 7.92 (1.04) & 44.44 (3.51) \\
& Conceptual & 26.04 (1.67) & 36.32 (1.85) & 42.96 (3.72) & 87.16 (5.46) \\
\midrule
\multirow{2}{*}{\textbf{TA}}
& GTA & 8.21 (0.99) & 19.87 (1.45) & 8.71 (1.37) & 51.44 (4.21) \\
& UTA & 37.63 (1.90) & 47.02 (1.95) & 44.40 (3.81) & 80.72 (4.97) \\
\bottomrule
\end{tabular}
\end{table}

\begin{table}[h]
\centering
\small
\caption{\revision{\textbf{Course-level summaries by subgroup for feedback stage time and feedback time.} Submissions in the treatment show longer feedback stage and feedback times across subgroups.  Values are means; standard errors (SEs) are shown in parentheses.}}
\label{tab:shared_stratified_2}
\begin{tabular}{llcccc}
\toprule
& & \multicolumn{2}{c}{\textbf{Feedback Stage Time Avg (sec)}} 
  & \multicolumn{2}{c}{\textbf{Feedback Time Avg (sec)}} \\
\cmidrule(lr){3-4} \cmidrule(lr){5-6}
\textbf{Category} & \textbf{Group} 
& \textbf{Control} & \textbf{Treatment} 
& \textbf{Control} & \textbf{Treatment} \\
\midrule
\multirow{2}{*}{\textbf{Mistakes}}
& No Mistakes & 2.83 (0.77) & 7.54 (0.73) & 6.43 (1.20) & 17.88 (2.34) \\
& Mistakes    & 7.73 (1.69) & 37.41 (17.65) & 9.48 (2.37) & 40.70 (20.16) \\
\midrule
\multirow{2}{*}{\textbf{Question}}
& Applied    & 3.68 (0.71) & 17.14 (8.05) & 10.98 (3.18) & 43.85 (27.37) \\
& Conceptual & 3.97 (1.20) & 11.54 (0.99) & 6.02 (0.83) & 22.29 (2.55) \\
\midrule
\multirow{2}{*}{\textbf{TA}}
& GTA & 3.39 (1.19) & 11.91 (0.96) & 6.80 (1.39) & 28.00 (3.07) \\
& UTA & 4.28 (0.73) & 16.81 (8.12) & 8.81 (1.98) & 34.91 (20.48) \\
\bottomrule
\end{tabular}
\end{table}

\begin{table}[h]
\centering
\small
\caption{\textbf{Course-level summaries by subgroup for adjusted feedback time and feedback ratings.} Adjusted feedback time and feedback rating remain broadly comparable across treatment and control conditions across subgroups. Values are means; standard errors (SEs) are shown in parentheses.}
\label{tab:shared_stratified_3}
\begin{tabular}{llcccc}
\toprule
& & \multicolumn{2}{c}{\textbf{\revision{Adjusted} Feedback Time Avg (sec/char)}} 
  & \multicolumn{2}{c}{\textbf{Feedback Rating Avg}} \\
\cmidrule(lr){3-4} \cmidrule(lr){5-6}
\textbf{Category} & \textbf{Group} 
& \textbf{Control} & \textbf{Treatment} 
& \textbf{Control} & \textbf{Treatment} \\
\midrule
\multirow{2}{*}{\textbf{Mistakes}}
& No Mistakes & 0.49 (0.11) & 0.23 (0.03) & 3.36 (0.06) & 3.37 (0.06) \\
& Mistakes    & 0.14 (0.03) & 0.58 (0.45) & 3.62 (0.09) & 3.64 (0.09) \\
\midrule
\multirow{2}{*}{\textbf{Question}}
& Applied    & 0.42 (0.10) & 0.81 (0.61) & 3.37 (0.08) & 3.42 (0.08) \\
& Conceptual & 0.13 (0.02) & 0.14 (0.02) & 3.54 (0.07) & 3.53 (0.07) \\
\midrule
\multirow{2}{*}{\textbf{TA}}
& GTA & 0.16 (0.03) & 0.15 (0.02) & 3.43 (0.07) & 3.57 (0.09) \\
& UTA & 0.30 (0.06) & 0.64 (0.45) & 3.50 (0.08) & 3.43 (0.07) \\
\bottomrule
\end{tabular}
\end{table}

\end{document}